\documentclass[11pt,twoside]{article}
\usepackage{amssymb,cite,graphicx}
\begin{document}
%
\makeatletter\@addtoreset{equation}{section}\makeatother
\def\theequation{\arabic{section}.\arabic{equation}}
%
	\def\ZZ{\mathbb{Z}}
        \def\NN{\mathbb{N}}
        \def\CC{\mathbb{C}}
        \def\RR{\mathbb{R}}
        \def\QQ{\mathbb{Q}}
        \def\PP{\mathbb{P}}
%
%
\def\adhoc{{\it ad hoc\/}}
\def\Adhoc{{\it Ad hoc\/}}
\def\ala{{\it \`a la\/}}
\def\ansatz{{\it ansatz\/}}
\def\ansatze{{\it ans\"atze\/}}
\def\Ansatze{{\it Ans\"atze\/}}
\def\apriori{{\it a priori\/}}
\def\Apriori{{\it A priori\/}}
\def\eg{\hbox{\it e.g.\/}}
\def\Eg{\hbox{\it E.g.\/}}
\def\etal{{\it et al.\/}}
\def\etc{{\it etc.\/}}
\def\ibid{{\it ibid.\/}}
\def\ie{\hbox{\it i.e.\/}}
\def\Ie{\hbox{\it i.e.\/}}

\def\mobius{m\"obius}
\def\Mobius{M\"obius}
\def\Kahler{K\"ahler}
\def\kahler{k\"ahler}

\newcommand{\beq}{\begin{equation}}
\newcommand{\eeq}{\end{equation}}
\newcommand{\beqa}{\begin{eqnarray}}
\newcommand{\eeqa}{\end{eqnarray}}
\newcommand{\N}{\mathbb{N}}
\newcommand{\Z}{\mathbb{Z}}
\newcommand{\Q}{\mathbb{Q}}
\newcommand{\R}{\mathbb{R}}
\newcommand{\C}{\mathbb{C}}
\newcommand{\e}{{\rm e}}
\newcommand{\bra}{\langle}
\newcommand{\ket}{\rangle}
\newcommand{\no}{\nonumber}
\newcommand{\hsp}{\hspace{5mm}}
\newcommand{\vsp}{\vspace{10mm}}
\newcommand{\tilQ}{\widetilde{Q}}
\newcommand{\Qn}{Q_0}
\newcommand{\tilQn}{\widetilde{Q}_0}
\newcommand{\M}{\mathcal{M}}
\newcommand{\MV}{{\mathcal{M}}_V}
\newcommand{\bM}{\overline{\mathcal{M}}}
\newcommand{\tilM}{\widetilde{\mathcal{M}}}
\newcommand{\mad}{\vec{m}_{\rm ad}}
\newcommand{\HH}{\mathcal{H}}
\newcommand{\z}{\zeta}
\newcommand{\tild}{\widetilde{d}}

%
%
\def\noj#1,#2,{{\bf #1} (19#2)\ }
\def\jou#1,#2,#3,{{\it #1\/ }{\bf #2} (19#3)\ }
\def\ann#1,#2,{{\it Ann.\ Physics\/ }{\bf #1} (19#2)\ }
\def\cmp#1,#2,{{\it Comm.\ Math.\ Phys.\/ }{\bf #1} (19#2)\ }
\def\ma#1,#2,{{\it Math.\ Ann.\/ }{\bf #1} (19#2)\ }
\def\ng#1,#2,{{\it Nagoya.\ Math.\ J.\/ }{\bf #1} (19#2)\ }
\def\jd#1,#2,{{\it J.\ Diff.\ Geom.\/ }{\bf #1} (19#2)\ }
\def\invm#1,#2,{{\it Invent.\ Math.\/ }{\bf #1} (19#2)\ }
\def\cq#1,#2,{{\it Class.\ Quantum Grav.\/ }{\bf #1} (19#2)\ }
\def\cqg#1,#2,{{\it Class.\ Quantum Grav.\/ }{\bf #1} (19#2)\ }
\def\ijmp#1,#2,{{\it Int.\ J.\ Mod.\ Phys.\/ }{\bf A#1} (19#2)\ }
\def\jmphy#1,#2,{{\it J.\ Geom.\ Phys.\/ }{\bf #1} (19#2)\ }
\def\jams#1,#2,{{\it J.\ Amer.\ Math.\ Soc.\/ }{\bf #1} (19#2)\ }
\def\grg#1,#2,{{\it Gen.\ Rel.\ Grav.\/ }{\bf #1} (19#2)\ }
\def\mpl#1,#2,{{\it Mod.\ Phys.\ Lett.\/ }{\bf A#1} (19#2)\ }
\def\nc#1,#2,{{\it Nuovo Cim.\/ }{\bf #1} (19#2)\ }
\def\np#1,#2,{{\it Nucl.\ Phys.\/ }{\bf B#1} (19#2)\ }
\def\pl#1,#2,{{\it Phys.\ Lett.\/ }{\bf B#1} (19#2)\ }
\def\pla#1,#2,{{\it Phys.\ Lett.\/ }{\bf A#1} (19#2)\ }
\def\pr#1,#2,{{\it Phys.\ Rev.\/ }{\bf #1} (19#2)\ }
\def\prd#1,#2,{{\it Phys.\ Rev.\/ }{\bf D#1} (19#2)\ }
\def\prl#1,#2,{{\it Phys.\ Rev.\ Lett.\/ }{\bf #1} (19#2)\ }
\def\prp#1,#2,{{\it Phys.\ Rept.\/ }{\bf C#1} (19#2)\ }
\def\ptp#1,#2,{{\it Prog.\ Theor.\ Phys.\/ }{\bf #1} (19#2)\ }
\def\ptpsup#1,#2,{{\it Prog.\ Theor.\ Phys.\/ Suppl.\/ }{\bf #1} (19#2)\ }
\def\rmp#1,#2,{{\it Rev.\ Mod.\ Phys.\/ }{\bf #1} (19#2)\ }
\def\yadfiz#1,#2,#3[#4,#5]{{\it Yad.\ Fiz.\/ }{\bf #1} (19#2) #3%
\ [{\it Sov.\ J.\ Nucl.\ Phys.\/ }{\bf #4} (19#2) #5]}
\def\zh#1,#2,#3[#4,#5]{{\it Zh.\ Exp.\ Theor.\ Fiz.\/ }{\bf #1} (19#2) #3%
\ [{\it Sov.\ Phys.\ JETP\/ }{\bf #4} (19#2) #5]}

%
%

\def\eq#1{.~(\ref{#1})}
\def\noeq#1{(\ref{#1})}
\hyphenation{eq}
\def\beq{\begin{equation}}
\def\eeq{\end{equation}}
\def\beqar{\begin{eqnarray}}
\def\eeqar{\end{eqnarray}}
\def\non{\nonumber}
\def\hE{\widehat{E}}

\newcommand{\be}{\begin{equation}}
\newcommand{\ee}{\end{equation}}
\newcommand{\bea}{\begin{eqnarray}}
\newcommand{\eea}{\end{eqnarray}}

\def\nfrac#1#2{{\displaystyle{\vphantom1\smash{\lower.5ex\hbox{\small$#1$}}%
        \over\vphantom1\smash{\raise.25ex\hbox{\small$#2$}}}}}

\def\u#1{{}^{#1}}
\def\d#1{{}_{#1}}
\def\p#1{\mskip#1mu}
\def\n#1{\mskip-#1mu}
\def\stop{\p6.}
\def\comma{\p6,}
\def\semi{\p6;}
\def\excl{\p6;}
\def\eqand{\p8 {\rm and}}
\def\eqor{\p8 {\rm or}}

%
%
\def\da{\downarrow}
\def\ua{\uparrow}
\def\upda{\updownarrow}
\def\to{\rightarrow}
\def\implies{\Rightarrow}
\def\To{\longrightarrow}
\def\longlongrightarrow{\relbar\joinrel\relbar\joinrel\rightarrow}
\def\ridiculousrightarrow{\relbar\joinrel\relbar\joinrel\relbar%
\joinrel\rightarrow}

\def\underarrow#1{\mathrel{\mathop{\longrightarrow}\limits_{#1}}}
\def\onnearrow#1{\mathrel{\mathop{\nearrow}\limits^{#1}}}
\def\undernearrow#1{\mathrel{\mathop{\nearrow}\limits_{#1}}}
\def\onarrow#1{\mathrel{\mathop{\longrightarrow}\limits^{#1}}}
\def\onArrow#1{\mathrel{\mathop{\longlongrightarrow}\limits^{#1}}}
\def\OnArrow#1{\mathrel{\mathop{\ridiculousrightarrow}\limits^{#1}}}
\def\lae{\mathrel{\mathop{\smash{\lower .5 ex \hbox{$\stackrel<\sim$}}}}}
\def\lae{\mathrel{\mathop{\smash{\lower .5 ex \hbox{$\stackrel>\sim$}}}}}
\def\eqq{\stackrel?=}


\def\ket#1{\left| #1 \right\rangle}
\def\bra#1{\left\langle #1 \right|}
\def\vev#1{\left\langle #1 \right\rangle}
\def\VEV#1{\left\langle #1 \right\rangle}
\def\f{\frac}
\def\pa{\partial}
\def\pb{\bar\pa}
\def\na{\nabla}
\def\Tr{{\rm Tr}}
\def\l:{\mathopen{:}\,}
\def\r:{\,\mathclose{:}}
\def\sech{\mathop{\rm sech}\nolimits}
\def\CF{{\cal F}}
\def\CL{{\cal L}}
\def\CM{{\cal M}}
\def\CT{{\cal T}}
\def\CR{{\cal R}}
\def\CY{{\cal Y}}
\def\mod{{\rm mod}}
\def\la{\langle}
\def\ra{\rangle}
\def\vr{\vec{r}}
\def\vq{\vec{q}}
\def\vs{\vec{s}}
\def\vm{\vec{m}}
\def\vom{\vec{\omega}}
\def\vp{\vec{\phi}}


%
%
%
%
%

%
%
%


\def\sltr{$SL(2,\RR)$}
\def\sltrouo{$SL(2,\RR)/U(1)$}
\def\half{\nfrac12}
\def\thrhalf{\nfrac32}
\def\d{{\rm d}}
\def\tdot{{\dot T}}
\def\fdot{{\dot f}}
\def\calo{{\cal O}}
\def\eipx{e^{\sqrt2 \p1 i \p1 p \p1 x}}
\def\eix#1{e^{\sqrt2 \p1 i \p1 #1 \p1 x}}
\def\ep#1{e^{#1 \sqrt2 \p1 \varphi}}
\def\epa#1#2{e^{#1 \sqrt2 \p1 #2 \p1 \varphi}}
\def\F{F\n4\left(\half, \half, 1, 1 - z\right)}
\def\Fsq{F^2\n4\left(\half, \half, 1, 1 - z\right)}
\def\Fd{F\n4\left(\half, \half, 2, 1 - z\right)}
\def\Fdsq{F^2\n4\left(\half, \half, 2, 1 - z\right)}
\def\Fp{F\n4\left(\half + |p|,\half + |p|,1 + |p|,1-z\right)}
\def\Fpd{F\n4\left(\half + |p|,\half + |p|,1 + |p|,1-z\right)}


\typeout{}
\typeout{}
\typeout{ M-Theory and  $N=1$ Moduli Space }
\typeout{}
\typeout{}
\typeout{Kentaro Hori, Hirosi Ooguri and Yaron Oz }
\typeout{}
\typeout{}
\typeout{THIS IS A LATEX FILE: LATEX TWICE, AS USUAL. }
\typeout{}
\typeout{}

\newcommand{\beqn}{\begin{equation}}
\newcommand{\eeqn}{\end{equation}}
\newcommand{\beqnarray}{\begin{eqnarray}}
\newcommand{\eeqnarray}{\end{eqnarray}}
\newcommand{\rd}{\partial}
\newcommand{\dfrac}[2]{ \frac{\displaystyle #1}{\displaystyle #2} }
\newcommand{\binom}[2]{ {#1 \choose #2} }
\newcommand{\res}{\;\mathop{\mbox{\rm res}}}
\newcommand{\cA}{{\cal A}}
\newcommand{\cAbar}{\bar{\cA}}
\newcommand{\cB}{{\cal B}}
\newcommand{\cBbar}{\bar{\cB}}
\newcommand{\cL}{{\cal L}}
\newcommand{\cLbar}{\bar{\cL}}
\newcommand{\cM}{{\cal M}}
\newcommand{\cMbar}{\bar{\cM}}
\newcommand{\cS}{{\cal S}}
\newcommand{\cSbar}{\bar{\cS}}
\newcommand{\cR}{{\cal R}}
\newcommand{\cO}{{\cal O}}
\newcommand{\cbar}{\bar{c}}
\newcommand{\tbar}{\bar{t}}
\newcommand{\ubar}{\bar{u}}
\newcommand{\vbar}{\bar{v}}
\newcommand{\Nbar}{\bar{N}}
\newcommand{\Wbar}{\bar{W}}
\newcommand{\lambdabar}{\bar{\lambda}}
\newcommand{\mubar}{\bar{\mu}}
\newcommand{\phibar}{\bar{\phi}}
\newcommand{\Phibar}{\bar{\Phi}}
\newcommand{\Psibar}{\bar{\Psi}}
\newcommand{\dbra}{\left<\!\left<}
\newcommand{\dket}{\right>\!\right>}
%

\newcommand {\nono} {\nonumber \\} 

\newcommand {\bear} [1] {\begin {array} {#1}}
\newcommand {\ear} {\end {array}}

\newcommand {\qeq} [1] {(\ref {eq:#1})}

\newcommand {\ul} [1] {\underline {#1}}

\newcommand {\myft} [2] 
        {\addtocounter {footnote} {#1}
         \footnotetext {#2}
         \addtocounter {footnote} {1}
        }

\newcommand {\myI} [1] {\int \! #1 \,}

\newcommand {\omy} {\mbox{$\Omega$}}

\newcommand {\tofl} {\mbox{$e^{i\phi_L}$}}
\newcommand {\tofr} {\mbox{$e^{i\phi_R}$}}
\newcommand{\CP}{\mathbb{C} {\rm P}}
\newcommand {\spr} {^{\prime}}

\newcommand {\beqarn} {\begin{eqnarray*}}
\newcommand {\eeqarn} {\end{eqnarray*}}

\newcommand {\sepe} {\;\;\;\;\;\;\;\;}
\newcommand     {\come} {\;\;\;\;}

\newcommand {\psibar} {\bar {\psi}}
\newcommand {\zbar} {{\bar z}}
\newcommand {\zb} {\zbar}
\newcommand {\CZ} {{\cal Z}}
\newcommand {\qbar} {{\bar q}}
\newcommand {\slZ} {{\mbox {$\mrm {SL} (2, \CZ)$}}}
\newcommand     {\abs}  [1] {{\left| #1 \right|}}
\newcommand {\brac} [1] {{\left\{       #1 \right\}}}
\newcommand     {\paren} [1] {{\left( #1 \right)}}
\newcommand     {\brak} [1] {{\left[ #1 \right]}}

\newcommand {\bs} {\backslash}
\newcommand {\rquo} [2] {\left. \bear {c} #1 
\\ \\ \ear \right/ \bear {c} \\ #2 \ear}
\newcommand {\lquo} [2] {\bear {c} \\ #1 \ear 
\left\backslash \bear {c} #2  \\ \\ \ear \right.}
\newcommand {\lrquo} [3] {\bear {c} \\ #1 \ear 
\left\backslash \bear {c} #2 \\ \\ \ear \right/
\bear {c} \\ #3 \ear}

\newcommand {\Mit} [1] {\mathit {#1}}
\newcommand {\mrm} [1] {\mathrm {#1}}

\newcommand {\im} {\mathrm {Im}}
\newcommand {\re} {\mathrm {Re}}
\newcommand {\tpa} {{\tilde \pa}}

\newcommand {\imply} {\Rightarrow}
\newcommand {\bij} {\leftrightarrow}
\newcommand {\bec} {\Leftarrow}

\newcommand {\nod} [1] {\mbox {$:#1\!:$}}

\newcommand {\comm} [2] {\mbox {$\left[ #1, #2 \right]$}}
\newcommand {\acomm} [2] {\mbox {$\left\{ #1, #2 \right\}$}}

\newcommand {\column} [1] {{\paren {\bear {c} #1 \ear}}}
\newcommand {\phys} {\ket {phys}}

\newcommand {\zNS}      {Z_{\mrm{NS}}}
\newcommand {\zR}       {Z_{\mrm{R}}}

\newcommand {\Fas} [1] {{F^{\brac {#1}}}}
\newcommand {\Aas} [1] {{A^{\brac {#1}}}}
\newcommand {\vol} {{\mbox {$\mrm{vol}$}}}
\newcommand {\tQ} {\tilde{Q}}

\leftline{
	\copyright~ 1997 International Press}
\leftline{
	Adv. Theor. Math. Phys. {\bf 1} (1997) \thepage--52}
\vspace{2cm}
\begin{center}
{\bf\LARGE
Strong Coupling Dynamics \\ 
of Four-Dimensional $N=1$ Gauge Theories \\
from M-Theory Fivebrane\\
}
\end{center}
\begin{center}
{\bf Kentaro Hori, Hirosi Ooguri, Yaron Oz}\\
\vspace{0.5cm}
Department of Physics,
University of California at Berkeley\\
366 Le\thinspace Conte Hall, Berkeley, CA 94720-7300, U.S.A.\\
and\\
Theoretical Physics Group, Mail Stop 50A--5101\\
Ernest Orlando Lawrence Berkeley National Laboratory\\
Berkeley, CA 94720, U.S.A.\\
\end{center}
\vspace{1cm}
\begin{abstract}
It has been known that the fivebrane of type IIA theory
can be used to give an exact low energy description 
of $N=2$ supersymmetric gauge theories in four dimensions.
We follow the recent M-theory description by Witten and 
show that it can be used to study theories with $N=1$ supersymmetry.
The $N=2$ supersymmetry can be broken to 
$N=1$ by turning on a mass for the adjoint chiral superfield in the 
$N=2$ vector multiplet. We construct the configuration of 
the fivebrane for both finite and infinite values 
of the adjoint mass. The fivebrane describes strong coupling 
dynamics of $N=1$ theory with $SU(N_c)$ gauge group and $N_f$ quarks.
For $N_c > N_f$, we show how the brane configuration 
encodes the information of the 
Affleck-Dine-Seiberg superpotential. For $N_c \leq N_f$, we study the 
deformation space of the brane configuration and compare it with the 
moduli space of the $N=1$ theory. We find agreement with field 
theory results, including the quantum deformation of the moduli 
space at $N_c=N_f$. We also prove the type II $s$-rule in M-theory
and find new non-renormalization theorems for $N=1$ superpotentials.
\end{abstract}

\section{Introduction}

In the past few years, we have learnt much about non-perturbative
dynamics of supersymmetric gauge theories and string theories.
In particular, the D(irichlet)-brane \cite{pol}
has provided an arena to exchange
results of gauge theories and string theories and to advance
our knowledge of both. This approach is very profitable since
the gauge coupling constant and the string coupling constant
are in general different. Therefore perturbative results in
one theory can be translated into non-perturbative statements
in the other theory. For example, if we compactify the type II
string on a singular Calabi-Yau three-fold and turn off the gravity,
we obtain an $N=2$ gauge theory in four dimensions. In this case,
the coupling constant of the gauge theory is some geometric
modulus of the three-fold, totally independent of the string coupling
constant \cite{kv,geom,sd}. Thus strong coupling dynamics of the gauge
theory
can be translated into facts about the geometry of the three-fold.

One can also obtain gauge theories in four dimensions by considering
webs of NS 5-branes and D4-branes in a
flat space in the type IIA string theory\footnote{This is
the T-dual of the configuration first introduced in \cite{hw}
to study aspects of $N=4$ gauge theories
in three dimensions. 
Configurations of intersecting branes
have also been used in order to count the microscopic
degrees of freedom of black holes
with various amounts of supersymmetry \cite{mal}.}.
A typical configuration consists of two parallel
NS 5-branes and several D4-branes suspended between them.
If world-volumes of the 5- and 4-branes share four flat dimensions,
we obtain $N=2$ gauge theory in four dimensions
with $SU(N_c)$ gauge group, where $N_c$ is the number
of \hbox{D4-branes}. We can add $N_f$ pairs of quark chiral multiplets
in the fundamental
representation of $SU(N_c)$ by attaching $N_f$ semi-infinite
D4-branes to one of the NS 5-branes.

In this construction, the distances between
the $N_c$ D4-branes suspended between the NS 5-branes
correspond to the vacuum expectation values (vevs)
of the adjoint scalar field
in the vector multiplet and therefore parametrize the Coulomb
branch of the model. If we turn on a mass of this adjoint
field, the $N=2$ supersymmetry is broken to $N=1$.
This corresponds to changing
the relative orientation of the two NS 5-branes while leaving their
common four dimensions intact \cite{bar}. If the NS 5-branes are
not parallel, the position of the D4-branes is fixed in order
to minimize their world-volume. Thus the Coulomb branch is lifted.
In the limit when the relative angle of the two NS 5-branes becomes
$\pi/2$, the adjoint mass becomes infinite and we obtain
the $N=1$ gauge theory with $SU(N_c)$ gauge group with $N_f$
quarks. This is the configuration studied in \cite{kut1,kut2}.

Recently Witten showed that one can give an exact
low energy description of the $N=2$ theory by reinterpreting
the brane configuration from the point of view of
M-theory \cite{witten}.  It is known that both
D4-branes and NS 5-branes of the type IIA theory come
from the fivebranes of M-theory, wrapped or unwrapped respectively
on the eleventh dimensional
circle with radius $R$. Thus the web of the D4-branes and the NS
5-branes
in the above may be considered as an $R \rightarrow 0$ limit
of a smooth configuration of a single fivebrane in M-theory.
The $N=2$ supersymmetry in four dimensions requires that the
world-volume of the fivebrane is ${\mathbb{R}}^{1,3} \times \Sigma$ and
that $\Sigma$ is holomorphically embedded in the ${\mathbb{R}}^3 \times S^1$
part of the eleven dimensions \cite{bbs} \footnote{Four-dimensional abelian
theory obtained from a fivebrane on ${\mathbb{R}}^{1,3} \times \Sigma$
was studied in \cite{ver}. A related observation was also made in 
\cite{1}.}.
Witten has shown
that, by imposing appropriate boundary conditions, the
configuration of $\Sigma$ is uniquely determined as
\beq
     t^2 - C_{N_c}(v, u_k)t + \Lambda_{N=2}^{2N_c-N_f}
                  \prod_{i=1}^{N_f} (v + m_i) = 0,
\label{curve}
\eeq
where $(v,t)$ are holomorphic coordinates on ${\mathbb{R}}^3 \times S^1$,
$C_{N_c}(v,u_k)$ is a polynomial in $v$ of degree $N_c$ with
coefficients which depend on the moduli $u_k$, and $m_i$
($i=1,...,N_f$) are the masses of the quarks.

In the M-theory description, the coupling constant of the
type IIA string is given by $g_s^2 \sim (R/l_{11})^3$
where $l_{11}$ is the
eleven-dimensional Planck scale, while the four-dimensional
gauge coupling constant scales as $g_{gauge}^2 \sim$ \break $R/L_{brane}$,
where $L_{brane}$ is a characteristic size of the brane
such as the distance between the NS 5-branes in the type IIA picture. 
Therefore we can take the limit, 
$R, L_{brane} \gg l_{11} \rightarrow 0$, 
while keeping the gauge coupling constant finite. 
Since the eleven-dimensional supergravity gives
the low energy description of M-theory, in this scaling limit
the configuration (\ref{curve}) should capture the strong coupling 
dynamics of the four-dimensional gauge theory. In fact, it turns 
out that $\Sigma$ given by (\ref{curve}) is the same as the Riemann surface
that appears in the exact solution to the $N=2$ gauge theory
\cite{sw,klty,af,us,apsp}.

Here we would like to make some historical remarks.
It has been pointed out earlier in \cite{sd}
that the Seiberg-Witten curve is geometrically realized as 
a configuration of the 5-brane. 
They considered the NS 5-brane in the type IIA string 
theory rather than the fivebrane of M-theory. However, 
at least in the case where there is $N=2$ supersymmetry 
in four dimensions, these two are essentially the same object. 

The purpose of this paper is to study the $N=1$ gauge theory, which
is obtained by turning on the adjoint mass in the $N=2$ theory,
using the fivebrane. We identify the fivebrane configuration
for non-zero value of the adjoint mass. As we mentioned before, 
the adjoint mass lifts the Coulomb branch of the
$N=2$ theory except at the roots of the Higgs
branches. Correspondingly we find that in order to rotate the
fivebrane we have to tune its moduli $u_k$ to completely degenerate
the curve $\Sigma$ and make it birationally equivalent to
$\CP^1$. We then show that there is a unique way to rotate
the fivebrane and determine how it is embedded in the eleven
dimensions.

 In the $N=1$ theory, a superpotential is non-perturbatively generated
for $N_c > N_f$ \cite{ads} and there are no supersymmetric vacua.
If we add quark masses and/or a perturbation quartic in the quark chiral
superfields that corresponds to the adjoint mass in the $N=2$ theory,
the theory has vacua characterized by the vevs
of $\tilde{Q}_i^a Q_a^j$, where $Q_a^i$, $\tilde{Q}_i^a$
are the squark fields \hbox{($i=1,...,N_f$, $a=1,...,N_c$)}.
In the type IIA picture, these vevs
are related to locations of the semi-infinite D4-brane.
We compute locations of the corresponding branches of
the M-theory fivebrane and show that they are in complete
agreement with the values of $\tilde{Q}_i^a Q_a^j$ obtained by
the field theory method, as functions of the quark and the adjoint
masses.

When $N_c \leq N_f$, there is an additional moduli space associated with
the  configuration of the fivebrane that corresponds
to a complete Higgsing of the gauge group. We study the structure of the
moduli space and compare it with that for the $N=1$ theory. We
also use the fivebrane to obtain non-renormalization theorems 
that have not been proven via the standard field theory method. 

This paper is organized as follows:

\noindent
Section 2 is devoted to field theory analysis.
We study the moduli space of vacua of
the $N=1$ theory which is obtained from the $N=2$
theory by adding a mass term to the adjoint chiral
multiplet. In particular we calculate the minima
of the $N=1$ superpotential which, for $N_c > N_f$,
consists of the Affleck-Dine-Seiberg superpotential,
a quartic perturbation in the quark chiral
superfields and a quark mass term. We prove a non-renormalization
theorem of such a superpotential and show that
its minima are precisely as expected if we start
with the $N=2$ theory and add a mass to the adjoint field.
We perform similar analysis when $N_f=N_c$ and
when $N_f \geq N_c+1$, where in the latter case we make use
of the magnetic dual theory.

In section 3, we develop techniques to study the moduli
space of vacua using the fivebrane of M-theory.
In particular we show in detail how  the Higgs branches of the
$N=2$ theory are described in this language.

In section 4, we show how to rotate the fivebrane to break
the $N=2$ supersymmetry to $N=1$ and present the resulting
configuration explicitly for non-zero value of the adjoint mass.
We find that the configuration encodes the strong coupling dynamics of
the $N=1$ theory. Specifically, we read vevs of the mesons  
 parametrizing the Higgs branches from the brane configuration
and show that they are in complete agreement
with the field theory results of section 2.

In section 5, we take the limit where the mass of
the adjoint is infinite. In the case of $N_c \leq N_f$,
we study the deformation space of the brane configuration
and compare it with the moduli space
of vacua of the $N=1$ theory.
Again we find a complete agreement with the field theory results.

In Section 6 we further study  the moduli space
of vacua of the $N=1$ theory. We find how the baryons are realized in the 
M-theory fivebrane and find again a complete agreement with the field theory
results of section 2. In this section we also prove the $s$-rule
of \cite{hw}. 

In section 7 we comment on the K\"ahler  potential in the M-theory 
framework.

We note that the method of intersecting branes
in IIA and IIB string theories and in some cases
their M-theory description have been applied
to the study of supersymmetric field theories in various dimensions in
\cite{2,3,4,5,6,7,8,9,10,11,12,13,14,15,16}

\section{Field Theory Analysis}

In this section we analyse in the field theory framework
the moduli space of vacua obtained by breaking $N=2$ to
$N=1$ by adding a mass term to the adjoint chiral multiplet.

\subsection{$N=2$ Moduli Space of Vacua}

We consider $N=2$ supersymmetric gauge theory with $SU(N_c)$ 
gauge group and $N_f$ quark
hypermultiplets in the fundamental representation.
In terms of $N=1$ superfields the vector multiplet  consists of
a field strength chiral multiplet $W_{\alpha}$ and a scalar
chiral multiplet $\Phi$,
both in the
adjoint representation of the gauge group. A quark hypermultiplet consists of
a chiral multiplet $Q$ in the $N_c$ and $\tilde{Q}$
in the $\bar{N_c}$ representation of the
gauge group.
The $N=2$ superpotential takes the form 
\beq
W = \sqrt{2} \tQ_i^a\Phi^b_aQ^i_b + \sqrt{2}m_j^i\tQ_i^aQ_a^j
\comma
\label{W}
\eeq
where $a,b=1,...,N_c; i,j=1,...,N_f$ and the quark mass matrix
$m={\rm diag}[m_1,...,m_{N_f}]$.

The $R$-symmetry group is $SU(2)_R\times U(1)_R$.
The bosons in the vector multiplet are singlets
under $SU(2)_R$  while the fermions in the vector multiplet form a doublet.
The fermions in the hypermultiplet are singlets under  
$SU(2)_R$  while the scalars in the  hypermultiplet form a doublet.
The theory is asymptotically free for $N_f < 2N_c$.
The instanton factor is proportional
to $\Lambda^{2N_c-N_f}$ where $\Lambda$ 
is the dynamically generated scale.
The $U(1)_R$ symmetry is anomalous and is broken to
$\Z_{2N_c-N_f}$.

The moduli space of vacua includes the Coulomb and Higgs branches. 
The Coulomb branch  is $N_c-1$ complex dimensional and
is parametrized by the gauge
invariant order parameters 
\beq
u_k = \langle\Tr(\phi^k) \rangle,~~~~~~~~~k=2,...,N_c
\comma
\label{gi}
\eeq
where $\phi$ is the scalar field in the vector multiplet.
Generically along the Coulomb branch the gauge group is broken
to $U(1)^{N_c-1}$.
The Coulomb branch structure is corrected by one loop effects and by 
instantons. The quantum Coulomb branch parametrizes a family of genus $N_c-1$
hyperelliptic curves
whose period matrix $\tau_{ij}$ is the low energy gauge coupling
\cite{sw,klty,af,us,apsp}.

Two types of  Higgs branches are distinguished \cite{aps}:
The baryonic branch and the non-
baryonic branches.
There is a single baryonic branch for $N_f \geq N_c$,
where generically the gauge
group is completely broken. Its complex dimension
is $2N_f N_c - 2(N_c^2-1)$.
The non-baryonic branches are classified by
an integer $r$ such that 
$1 \leq r \leq \min \{[N_f/2], N_c-2 \}$. The \hbox{$r$-th}
non-baryonic branch has complex dimension 
$2r(N_f-r)$.
The baryonic branch emanates
from a point in the Coulomb
branch while
the non-baryonic
branches
emanate from submanifolds in the Coulomb branch
(of dimension $N_c-r-1$ for the $r$-th non-baryonic branch)
and constitute mixed branches.
The Higgs branches are determined classically; however,
where they intersect with each other
and with the Coulomb branch is modified quantum mechanically.

\subsection{Breaking $N=2$ to $N=1$}

The $N=2$ supersymmetry is broken to $N=1$  by turning on
a bare mass $\mu$ for the adjoint
chiral multiplet $\Phi$
\beq
W = \sqrt{2} \Tr(\tilQ\Phi Q) + \mu \Tr(\Phi^2)
\stop
\label{Wp}
\eeq
When the mass for the adjoint chiral multiplet is small
we can still use the low energy description of \cite{sw},
and it turns out that
the structure of the moduli space 
of vacua is modified as follows.
Most of the Coulomb branch is lifted except for a discrete set of points.
$2N_c-N_f$ points related to each other by the action of $\Z_{2N_c-N_f}$
and which correspond to to points in the moduli space of vacua
where all the $\alpha$ cycles of the hyper-elliptic curve vanish remain.
The root of the baryonic branch as well as the baryonic branch itself remain.
The non-baryonic branches remain but instead of
being mixed branches they emanate from points.
More precisely, the $r$-th non-baryonic branch that 
emanated from
a submanifold of dimension $N_c-r-1$ in the Coulomb branch
is now emanating from $2N_c-N_f$ points related
to each other by  $\Z_{2N_c-N_f}$
with the exception of the
$r=N_f/2$ case ($N_f$ even) where
the $\Z_2$ subgroup is unbroken
and the $\Z_{2N_c-N_f}$ orbit consists of $N_c-N_f/2$ points.

When the mass $\mu$ for the adjoint
chiral multiplet is increased beyond $\Lambda_{N=2}$,
the renormalization group flow below the scale $\mu$
is the same as in $N=1$ Supersymmetric QCD (SQCD) 
with the dynamical scale $\Lambda_{N=1}$ given by
\beq
\Lambda_{N=1}^{3N_c-N_f} = \mu^{N_c}\Lambda_{N=2}^{2N_c-N_f}
\,.
\label{RG}
\eeq
If $\mu$ is much larger than $\Lambda_{N=1}$ but finite,
we can first integrate out the heavy field $\Phi$ in (\ref{Wp}), 
obtaining a superpotential which is
quartic in the quark chiral superfields
and is proportional to $1/\mu$:
\beq
\mathit{\Delta W} = \frac{1}{2\mu}
\left(\Tr(M^2) - \frac{1}{N_c}(\Tr M)^2\right),
\label{DW}
\eeq
where $M=\tilQ Q$.
In other words, we may consider the system below the energy scale $\mu$
as the $N=1$ SQCD with
the tree level superpotential $\mathit{\Delta}W$ and the
dynamical scale $\Lambda_{N=1}$ given by (\ref{RG}).
As we send the mass $\mu$ to infinity keeping $\Lambda_{N=1}$
finite, 
the potential $\mathit{\Delta}W$ disappears and the system 
becomes equivalent to $N=1$ SQCD, whose
low energy properties depend on the number of flavors $N_f$.
The structure of the moduli space of vacua should match
for finite values of $\mu$ with
the one that we get by starting with $N=2$ and adding mass
for the adjoint chiral multiplet.
This will be checked in the following.

\bigskip
\noindent
\underline{Pure Yang-Mills Theory $N_f=0$}

\medskip
For $N=1$ $SU(N_c)$ Yang-Mills theory,
there are $N_c$ massive vacua where the discrete $\Z_{2N_c}$
$R$-symmetry is spontaneously broken to $\Z_2$,
as the computation of the Witten index \cite{Windex} shows.
They correspond to $N_c$ curves in the $N=2$ theory with all
the $\alpha$ cycles vanishing.
These curves are related by the action of the discrete $\Z_{2N_c}$
$R$-symmetry group, which is consistent with the structure of the
$N=1$ vacua.

\bigskip
\noindent
\underline{$0<N_f<N_c$}

\medskip
In $N=1$ SQCD with the the number of flavors in this region,
a superpotential is
dynamically generated \cite{ads}: for $N_f=N_c-1$ it is due to the
effect of instantons, and for the other cases it is
due to a strong gauge dynamics. It takes the form
\beq
W_{\rm ADS} =  (N_c-N_f)\left(\frac{\Lambda_{N=1}^{3N_c-N_f}}{\det M}
\right)^{1/(N_c-N_f)} \,,
\label{ADS}
\eeq
and thus there is no supersymmetric vacuum.

For large but finite $\mu$,
at the scale far below $\mu$ but much larger
than $\Lambda_{N=1}$,
the quartic term is very small and can be considered as
a perturbation to the ordinary $N=1$ system.
Therefore the superpotential (\ref{ADS}) is generated
in this case as well.
Thus, we expect that the effective superpotential is just the sum
\beq
\begin{array}{rl}
W_{\it eff} &=  (N_c-N_f)\left(\frac{\Lambda_{N=1}^{3N_c-N_f}}{\det M}
\right)^{1/(N_c-N_f)} + \\
&+\frac{1}{2\mu}
\left(\Tr(M^2) - \frac{1}{N_c}(\Tr M)^2\right)
\,.
\end{array}
\label{WM}
\eeq
In fact, this is an {\it exact} superpotential which is valid for
any non-zero value of $\mu$.
This follows from the following holomorphy argument \cite{is}.
The superpotential must be an analytic function around
the decoupling limit $1/\mu=0$ and therefore can be expanded
with respect to $1/\mu$
where the first two terms are fixed to be (\ref{WM}).
Thus, a term that can be generated takes the form
\beq
\mu^{-\alpha}M^{\beta}\Lambda_{N=1}^{(3N_c-N_f)\gamma},
\label{can}
\eeq
where $\alpha$ is a non-negative integer and
$M^{\beta}$ is some
combination of the meson matrix of order $\beta$ which is invariant
under the flavor group $SU(N_f)$.
We require
$\gamma\geq 0$ for the existence of the weak coupling limit
$\Lambda_{N=1}\to 0$.
We recall that
$\mu$, $M$, and $\Lambda^{3N_c-N_f}$ carry the following
$U(1)_R\times U(1)_A$ charges where $U(1)_R$ is the anomaly free
combination of the $U(1)$ $R$-symmetry group, while $U(1)_A$
is the axial flavor symmetry which is anomalously broken.
\beq
\begin{array}{cccc}
&\Lambda_{N=1}^{3N_c-N_f}&M&\mu\\[0.2cm]
U(1)_R&0&2{N_f-N_c\over N_f}&2{N_f-2N_c\over N_f}\\
U(1)_A&2N_f&2&4\\
\end{array}
~~ .
\eeq
The charges of $\mu$ are determined so that the perturbation
term $\int \d^2\theta
\mathit{\Delta}W$ is invariant, and the $U(1)_A$ charge of
the instanton factor
$\Lambda_{N=1}^{3N_c-N_f}$ reflects the axial anomaly.
The perturbation term by
(\ref{can}) must be invariant under $U(1)_R\times U(1)_A$
with this assignment of the charges, and this requires
$N_f(-\alpha+\beta-1)=N_c(-2\alpha+\beta)$ and
$-2\alpha+\beta=-N_f\gamma$, and thus in particular
\beq
1-\alpha=(N_c-N_f)\gamma.
\eeq
Since we are considering the case $N_f<N_c$, this together with
$\gamma\geq 0$ requires $\alpha=0$,$\gamma=1/(N_c-N_f)$
or $\alpha=1$, $\gamma=0$. The former corresponds to the
Affleck-Dine-Seiberg potential (\ref{ADS}), and
the latter corresponds to the tree level term $\mathit{\Delta}W$.
In this way, we have seen that the superpotential (\ref{WM})
is exact.

The moduli space of vacua is the variety of extrema of this
superpotential. We now determine this.
Extremizing (\ref{WM}) we have
\beq
M^2 -   \frac{1}{N_c}(\Tr M)M = \mu
\left(\frac{\Lambda_{N=1}^{3N_c-N_f}}{\det M} \right)^{1/(N_c-N_f)} 
\stop
\label{min}
\eeq
Let us  perform a similarity transformation
$M \rightarrow M' = g M g^{-1}$ in (\ref{min}) 
such that 
$M'_{i+j,i} = 0, j> 0$ and define $m_i:=M_{ii}$.
It is not possible for more than two of the elements
$m_i$ to have different values.
In order to see this
 suppose that there are three different values $m_i, m_j,m_k$.
Equation (\ref{min}) for the diagonal elements 
\beq
 m_i^2 -   \frac{1}{N_c}(\sum_{l=1}^{N_f}m_l) m_i  = \mu
 \left(\frac{\Lambda_{N=1}^{3N_c-N_f}}{\prod_l m_i} \right)^{1/(N_c-N_f)} 
 \label{diag}
 \eeq
implies upon subtracting the equations for any two of $m_i, m_j,m_k$, that
\beq
m_i+m_j = m_j+m_k = m_i + m_k
\comma
\eeq
in contradiction with the assumption that they all have different values.

In fact  $M$ can be diagonalized. In order to see that
assume that this is not the case.
Then $M$ can be brought to the Jordan form.
In this case we will get from off diagonal entries in (\ref{min}) 
\beq
2m_1 -\frac{1}{N_c}\Tr M = 0
\stop
\label{eq}
\eeq
If we assume that there exists another diagonal entry value
$m_2 \neq m_1$ we will
also get an equation of the form
\beq
m_1+m_2 -\frac{1}{N_c}\Tr M = 0
\comma
\eeq
in contradiction with (\ref{eq}).
The Jordan form implies that all the diagonal entries must be the same
and therefore (\ref{eq}) yields $2m_1 = m_1 N_f/N_c$,
which cannot be satisfied since
$N_f < N_c$.
We are led to a contradiction and thus $M$ can be diagonalized.

Since there are no more than two possible values for
the diagonal entries we find 
two types of solutions to (\ref{diag}).
In the first type  all the diagonal entries  are equal,
$m_1=...=m_{N_f} = m$, where
\beq
m = \left(\frac{N_c}{N_c-N_f}\right )^{\frac{N_c-N_f}{2N_c-N_f}}
\mu \Lambda_{N=2}
\comma
\label{r0}
\eeq
and we used the RG matching relation (\ref{RG}).
This solution corresponds  
 in the $N=2$ picture to the case when all the $\alpha$ cycles
 degenerate,
and will be 
 denoted as the $r=0$ case.
In the second type there are two different entries on the diagonal.
They can be made to take the form
$m_1=...=m_r = m_{(1)} \neq m_{r+1}=...=m_{N_f} = m_{(2)}$.
The equations determining $m_{(1)}$ and $m_{(2)}$ are
\beqa
&&\left(1-{r\over N_c}\right)\,m_{(1)}
+\left(1-{N_f-r\over N_c}\right)\,m_{(2)}
=0\,,
\label{la}\\[0.2cm]
&&m_{(1)}^{N_c-N_f+r}m_{(2)}^{N_c-r}
=(-1)^{N_c-N_f}(\mu \Lambda_{N=2})^{2N_c-N_f}
\label{laa}
\eeqa
and the solution is
\beq
m_{(2)} \,=\, \left((-1)^r
\left(\frac{N_c-r}{N_c - N_f + r}\right )^{N_c-N_f+r} \right )^{\frac{1}
{2N_c-N_f}} \mu \Lambda_{N=2}
\stop
\label{sol}
\eeq
where $m_{(1)}$ is given by (\ref{la}).

To summarize:
The solutions (\ref{sol}) are classified by an
integer $r$ that takes values in the
range $ 0 \leq r \leq [N_f/2]$.
The moduli space consists of the orbits of the complexified
flavor group
$GL(N_f,\C)$ through such diagonal solutions.
Since the diagonal solution for $r$ is invariant under the subgroup
$GL(r,\C)\times GL(N_f-r,\C)$,
the moduli space is the homogeneous space
$GL(N_f,\C)/(GL(r,\C)\times GL(N_f-r,\C))$, which has complex dimension
$2r(N_f-r)$.
This dimension agrees with what we
expect from the $N=2$ discussion where $r$ is the parameter that
characterizes the 
non-baryonic branches.      
As we see from the solution (\ref{sol}), for each $r<N_f/2$
there are $2N_c-N_f$ solutions
related by the action of $\Z_{2N_c-N_f}$, as expected.
For $r=N_f/2$ ($N_f$ even), however,
since $m_{(1)}=-m_{(2)}$, the two solutions related by the $\Z_2$
subgroup (sign change) are related by conjugation by an element
of $GL(N_f,\C)$, and thus are in the same orbit.
So there are only $N_c-N_f/2$ families, which is also expected.

In the limit $\mu\to\infty$ keeping $\Lambda_{N=1}$ finite,
all these solutions diverge since
$(\mu \Lambda_{N=2})^{2N_c-N_f}=\mu^{N_c-N_f}\Lambda_{N=1}^{3N_c-N_f}$.
This is consistent with the fact that there is no supersymmetric
vacuum for the $N=1$ SQCD in this region of the flavor.

\bigskip
{\it Inclusion of Bare Mass}

\medskip
Let us consider the case where the quark mass term
$\sum (m_f)_{i,j}\tilQ^i Q_j$ 
is turned on. The effective superpotential is given by
\beq
W_{\it eff}=W_{\rm ADS}+\mathit{\Delta}W+\Tr(m_fM)\,,
\label{WMm}
\eeq
where $W_{\rm ADS}$ and $\mathit{\Delta}W$ are given by (\ref{ADS})
and (\ref{DW}). Again, this is an exact superpotential as can
be seen from the analyticity at $m_f=1/\mu=0$ and charge conservation,
where $m_f$ carries $U(1)_R\times U(1)_A$ charge
$(2N_c/N_f, -2)$.

Extremizing  $W_{\it eff}$, we obtain the moduli space of vacua.
Here we present the result in the case where the mass matrix
is proportional to the identity matrix, $(m_f)_{i,j}=m_f\delta_{i,j}$.
In this case, as in the previous discussion,
$M$ is diagonalizable\footnote{
For some special values of $m_f$ a Jordan block of size two is
allowed.},
and there are at most two kinds
of eigenvalues. Thus, it is again classified by
$r=0,1,\ldots,[N_f/2]$.
The equations determining the two (or one in the case $r=0$)
eigenvalues $m_1=\cdots =m_r=m_{(1)}$ and
$m_{r+1}=\cdots =m_{N_f}=m_{(2)}$ are
in this case
\beqa
&&\left(1-{r\over N_c}\right)\,m_{(1)}
+\left(1-{N_f-r\over N_c}\right)\,m_{(2)}
+\mu m_f=0\,,
\label{1st}\\[0.2cm]
&&m_{(1)}^{N_c-N_f+r}m_{(2)}^{N_c-r}
=(-1)^{N_c-N_f}(\mu \Lambda_{N=2})^{2N_c-N_f}\,.
\label{2nd}
\eeqa
Note that there are $2N_c-N_f$ solutions for each $r$, but
they are not related any longer by the discrete $R$-symmetry group
$\Z_{2N_c-N_f}$ which is explicitly broken by the quark mass term.

Lets us consider the limit $\mu\to \infty$ keeping
$\Lambda_{N=1}$ finite
where the system becomes the $N=1$ SQCD with massive quarks.
Since
$(\mu \Lambda_{N=2})^{2N_c-N_f}=\mu^{N_c-N_f}\Lambda_{N=1}^{3N_c-N_f}$,
$m_{(1)}$ or $m_{(2)}$ must diverge in the limit.
Because a vacuum must have a finite vev for $M=\tilQ Q$,
only the solutions with $r=0$ and
$m_{(2)}$ finite remain as supersymmetric vacua.
In this case, $m_{(1)}$ diverges as
$m_{(1)}\sim -\mu m_f$ as follows from the first equation (\ref{1st}).
Then, inserting this to the second equation, we see that
$M^i_j=m_{(2)}\delta^i_j$ where
\beq
m_{(2)}^{N_c}=\Lambda_{N=1}^{3N_c-N_f}/m_f^{N_c-N_f},
\eeq
in the limit $\mu\to\infty$. Thus, we have $N_c$ vacua.
This is consistent with the interpretation of the
low energy physics as the pure $N=1$ Yang-Mills theory.

\bigskip
\noindent
\underline{$N_f=N_c$}

\medskip
In $N=1$ SQCD with this number of flavors,
the classical moduli space of vacua is modified quantum mechanically.
It is parametrized by the meson $M=\tilde{Q}Q$ and the baryons $B=Q^{N_c},
\tilde{B}= \tilde{Q}^{N_c}$ satisfying the constraint \cite{seiberg}
\beq
\det M - \tilde{B}B = \Lambda_{N=1}^{2N_c}
\stop
\label{bm}
\eeq

As  in the case $N_f < N_c$, for large finite $\mu$,
at the scale far below $\mu$ and much larger
than $\Lambda_{N=1}$,
the quartic term (\ref{DW})  is very small and can be considered as
a perturbation to the ordinary $N=1$ system.
We expect that the effective superpotential is   
\beq
W_{\it eff} =  X(\det M - \tilde{B}B - \Lambda_{N=1}^{2N_c})
 + \frac{1}{2\mu}
\left(\Tr(M^2) - \frac{1}{N_c}(\Tr M)^2\right)
\, ,
\label{sup}
\eeq
where we introduced a Lagrange multiplier $X$ to
impose the constraint (\ref{bm}).
In this case the holomorphy and global symmetries are not powerful enough
to ensure that (\ref{sup}) is the exact superpotential.
In section 4, we will see by brane analysis that
this is indeed the case.

Extremizing  $W_{\it eff}$ we get equation (\ref{bm}) from the derivative
with respect to $X$, 
\beq
X B = 0,~~~~~~ X \tilde{B} = 0
\comma
\label{X}
\eeq
from the derivative
with respect to $\tilde{B},B$ and
\beq
M^2 -   \frac{1}{N_c}(\Tr M)M = -\mu X \det M  
\comma
\label{mini}
\eeq
from the  derivative
with respect to $M$.

Consider the last equation. Similar analysis as
the one done in the $N_f < N_c$ case shows
that $M$ can be diagonalized and can have at most
two different eigenvalues, which as before
we denote by  $m_1=...=m_r = m_{(1)} \neq m_{r+1}=...=m_{N_f} = m_{(2)}$.
There are two cases to consider, $X=0$ and $X\neq 0$.
When $X=0$ then for $\tilde{B}B=0$ we get using (\ref{bm})
 that the solution to (\ref{mini}) is given by
\beq
m_{(1)} = m_{(2)} = \Lambda_{N=1}^2
\comma
\label{m12}
\eeq
while for  $\tilde{B}B\neq 0$ we have $m_{(1)} = m_{(2)} =m$ related to 
$\tilde{B}B$ by 
\beq
m^{N_c} -  \tilde{B}B = \Lambda_{N=1}^{2N_c}   
\stop
\label{mbb}
\eeq

When $X\neq 0$ 
the equations determining $m_{(1)}$ and $m_{(2)}$ are
\beqa
&&\left(1-{r\over N_c}\right)\,m_{(1)}
+\left(1-{N_c-r\over N_c}\right)\,m_{(2)}
=0\,,
\label{la1}\\[0.2cm]
&&m_{(1)}^{r}m_{(2)}^{N_c-r}
=(\mu \Lambda_{N=2})^{N_c}
\label{la2}
\eeqa
and the solution to (\ref{la2}) is given by
\beqa
     m_{(1)} &=& \left( (-1)^{N_c-r} 
          \left( \frac{r}{N_c-r} \right)^{N_c-r}
          \right)^{\frac{1}{N_c}}\mu\Lambda_{N=2}\nonumber \\
& & \\
     m_{(2)}&=& \left( (-1)^r 
          \left( \frac{N_c-r}{r} \right)^{r}
          \right)^{\frac{1}{N_c}}\mu \Lambda_{N=2}\nonumber
\label{r}          
\eeqa
and $B=\tilde{B}=0$ by (\ref{X}). 

As in the $N_f < N_c$ case, the solutions (\ref{r})
correspond to the $r$-th non-baryonic
branch of complex dimension $r$. 
As we see from the solution (\ref{r}), for each $r$
there are $N_c$ solutions
related by the action of $\Z_{N_c}$. 
The solution (\ref{mbb}) corresponds to
a new branch which did not exist in the $N_f <N_c$ 
region. This is a complex dimension one submanifold
of the baryonic branch whose complex dimension is two. This branch
corresponds to a complete Higgsing of the gauge group.
 
In the limit $\mu\to\infty$ keeping $\Lambda_{N=1}^2$ finite,
all these solutions remain with 
$\mu \Lambda_{N=2}=\Lambda_{N=1}^2$ and they define submanifolds of the 
$N_c^2+1$ complex dimensional  Higgs branch of $N=1$ SQCD.

\bigskip
\noindent
\underline{$N_c+1\leq N_f$}

In this case the classical space of vacua of $N=1$ SQCD is 
not modified quantum mechanically.
It is parametrized by the mesons and baryons which
are related by a classical constraint.
However,
it is useful in this case to  use the dual magnetic description based on 
the gauge group $SU(N_f-N_c)$ with $N_f$ flavors of quarks  $q, \tilde{q}$ 
and gauge invariant fields $M$ with the superpotential
\cite{seiberg2}
\beq
W_{mag}= \frac{1}{\lambda}\tilde{q}Mq +
(N_c-N_f)\left(\frac{\Lambda_{N=1}^{3N_c-N_f}}{\det M}
\right)^{1/(N_c-N_f)}
\stop
\label{mag}
\eeq
The scale $\lambda$ relates the scale $\Lambda_{N=1}$ of the electric
theory and the scale   $\tilde{\Lambda}_{N=1}$of the magnetic theory 
by
\beq
\Lambda_{N=1}^{3N_c-N_f}\tilde{\Lambda}_{N=1}^{3(N_f-N_c)-N_f}
=(-1)^{N_f-N_c} \lambda^{N_f}
\comma
\eeq
and is used in order to relate the electric 
and magnetic gauge invariant operators.
The dimension two (at the UV fixed point)
 meson $M$ in the electric description is related to
the dimension one singlets $M_{mag}$
 in the electric description by $M=\lambda M_{mag}$.
 Similarly,
 the baryons of the electric theory $B,\tilde{B}$ are related to the 
baryons of the magnetic theory $B_{mag},\tilde{B}_{mag}$
constructed from the dual quarks
by $B=C B_{mag}, \tilde{B} = C \tilde{B}_{mag}$ with 
$C=\left(-(-\lambda)^{N_c-N_f}\Lambda_{N=1}^{3N_c-N_f}\right)^{1/2}$.
 Equation (\ref{mag}) is written
 in terms of 
 the electric meson and the scale of the electric theory. 

Consider  the effective superpotential  
\beq
W_{\it eff} = W_{mag}
 + \frac{1}{2\mu}
\left(\Tr(M^2) - \frac{1}{N_c}(\Tr M)^2\right)
\stop
\label{sup1}
\eeq
As in the case $N_f=N_c$ we do not have a field theory proof
that the superpotential
(\ref{sup1}) is exact. The brane picture in section 6 suggests that it is.
Let us set the vev of the magnetic quarks to zero. This corresponds
 to setting the 
vev of the baryon operator $B,\tilde{B}$
to zero and studying the non-baryonic branches.
The superpotential that we get in this case is identical
to the one analysed in the 
region $N_f < N_c$.
Therefore, its extrema are precisely those given by equations
(\ref{r0}), (\ref{la}),
(\ref{laa}) and
(\ref{sol}).

Consider now the case where
the vev of the magnetic quarks
in (\ref{sup1}) is different from zero. 
We will not study the complete moduli space of vacua obtained as the extrema
of (\ref{sup1}) and restrict ourselves to those
solutions which will be needed in section 6
for comparison with the M-theory fivebrane.
We start with  $N_f=N_c+1$. In this case
$W_{mag}$ does not contain a fractional power
of ${\det M}$ and  it is straightforward to see that 
\beq
M = {\rm diag}[0,m,...,m],~~~~~B_{mag} = (b,0,...,0),~~~~~
\tilde{B}_{mag}^{t} = (\tilde{b},0,...,0)
\comma
\label{MBB}
\eeq
is a solution to the extrema of (\ref{sup1}) when
\beq
\frac{m^{N_c}}{\Lambda_{N=1}^{3N_c-N_f}} - \frac{\tilde{b}b}{\lambda}
 = \frac{m}{\mu}
\label{ncnf1}
\stop
\eeq
In terms of the gauge invariant operators of the electric theory
(\ref{ncnf1}) reads
 \beq
m^{N_c} - \tilde{B}B
 = \frac{m}{\mu}\Lambda_{N=1}^{3N_c-N_f}
\label{ncf}
\stop
\eeq

Consider next the region $N_f > N_c+1$, and  the following generalization 
of (\ref{MBB})
\beq
M = {\rm diag}[\overbrace{0,..,0}^{N_f-N_c},\overbrace{m,...,m}^{N_c}],
~~~~~q_{ij} = q_i \delta_{ij}~~~~~
\tilde{q}_{ij} = \tilde{q}_i\delta_{ij} 
\comma
\eeq
where we recall that the magnetic quarks
$q,\tilde{q}$ are $N_f\times (N_f-N_c)$ and
 \hbox{$(N_f-N_c) \times N_f$} matrices 
 respectively.

In order to find the extrema of
 (\ref{sup1}) we have to specify how to take a
limit to approach the region ${\det M}=0$.
In fact, there is an ambiguity in defining the limit.
One way to take the limit is to set \hbox{$M_{kk}=\varepsilon
e^{-\frac{2 \pi i k}{N_f-N_c}}$}, \hbox{$k=1,...,N_f-N_c$}, and take
 $\varepsilon$ to zero.
 Extremizing (\ref{sup1}) we obtain 
\beq
e^{\frac{2 \pi i k}{N_f-N_c}}\left(\frac{m^{N_c}}
{\Lambda_{N=1}^{3N_c-N_f}}\right)^{\frac{1}{N_f-N_c}} - 
\frac{\tilde{q}_kq_k}{\lambda}
 = \frac{m}{\mu},~~~~~~k=1,...,N_f-N_c
\label{nfc}
\stop
\eeq
Taking the product of equations (\ref{nfc}) and using the gauge 
 gauge invariant operators of the electric theory, this becomes
 \beq
m^{N_c} - \tilde{B}B
 = {\mu}^{N_c-N_f} m^{N_f-N_c}\Lambda_{N=1}^{3N_c-N_f}
\label{final}
\stop
\eeq
If we approach ${\det M}=0$ in a different direction, we obtain
a different relation between $m$ and $\tilde{B}B$. As we see later, 
there is no ambiguity of this type in the fivebrane 
description. In fact, the fivebrane chooses this particular way
to take the limit. We will discuss more on this issue
in section 6.

We close this section with several comments.
 The baryonic branch which is the branch that
 includes non zero vev for the baryon operator $B,\tilde{B}$
  has complex dimension 
 $2N_fN_c -2(N_c^2-1)$ for finite $\mu$, and we explored only part of it.
In the limit $\mu\to\infty$ keeping $\Lambda_{N=1}$ finite,
the solutions 
(\ref{r0}), (\ref{la}) and
(\ref{sol}) vanish since
$(\mu \Lambda_{N=2})^{2N_c-N_f}=\mu^{N_c-N_f}\Lambda_{N=1}^{3N_c-N_f}$.
This is consistent with the fact that on the moduli space
of vacua of $N=1$ SQCD with $N_f \geq N_c+1$,
${\det M}=0$ when $\tilde{B}B=0$.
The solutions (\ref{ncf}) and (\ref{final}) in this limit describe part of
the moduli space
of vacua of $N=1$ SQCD where $\tilde{B}B\neq 0$. 
In the limit $\mu\to\infty$ extra $N_c^2-1$ complex degrees of freedom
become massless and 
the Higgs branch of the theory is $2N_cN_f -(N_c^2-1)$
complex dimensional with the baryonic
and non-baryonic branches being submanifolds of it.

\medskip

\section{$N=2$ Higgs Branch via M-Theory Fivebranes}

In this section we analyse
the moduli space of vacua of $N=2$ SQCD by using fivebranes in
M-theory. In particular, we study how the Higgs branch 
of the system is geometrically realized in this picture. This
provides the starting point for the studies presented in
the following sections.

\begin{figure}[htb]
\begin{center}
\includegraphics[width=4.2in]{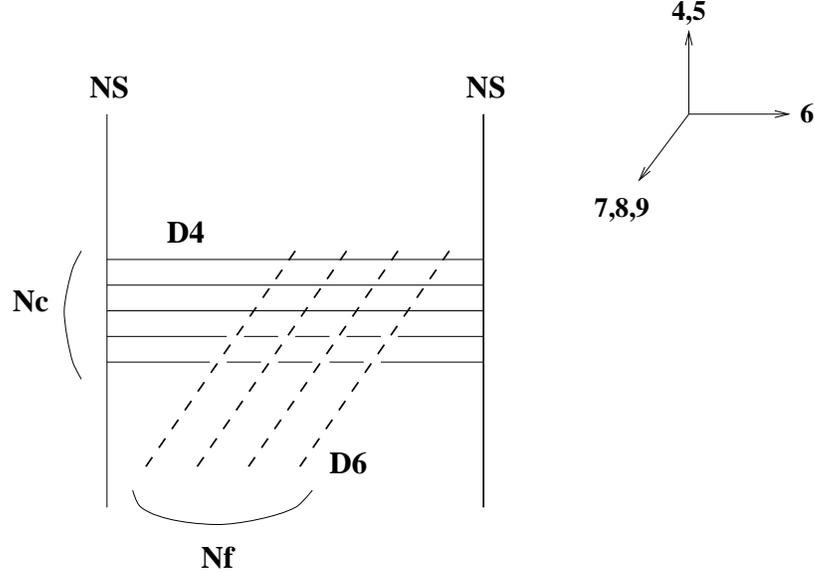}
\end{center}
\caption{The type IIA picture of
$N=2$ supersymmetric $SU(N_c)$ 
gauge theory with $N_f$ flavors (Coulomb branch).}
\label{fig1}
\end{figure}

Let us first review the description of the Higgs branch
in the type IIA picture. 
Consider the brane configuration of \cite{witten} that
preserves eight supercharges 
in type IIA string theory on a flat space-time
with time $x^0$ and space coordinates $x^1,\ldots,x^{9}$.
The brane configuration depicted in figure \ref{fig1} consists of
two NS 5-branes with worldvolume
coordinates $x^0,x^1,x^2,x^3,x^4,x^5$, $N_c$ D4-branes
suspended between them
with  worldvolume
coordinates $x^0,x^1,x^2,\\ x^3,x^6$ and $N_f$ D6-branes with 
  worldvolume
coordinates $x^0,x^1,x^2,x^3,x^7,x^8,\\ x^9$.
Since the D4-brane is finite in the $x^6$ direction, the low energy
effective theory on the web of branes is the four-dimensional
$N=2$ supersymmetric gauge theory
on its worldvolume coordinates  $x^0,x^1,x^2,x^3$.
The theory has $SU(N_c)$ gauge group and $N_f$ hypermultiplets
in the fundamental representation
of the gauge group.

Figure 1 depicts the Coulomb branch of the theory. To go to
the Higgs branch, we break the D4-branes on the D6-branes
and have them suspended between the D6-branes. Motion 
of the D4-branes along the D6-branes describes the Higgs branch.    
The location of a D4-brane between two D6-branes is parametrized 
by two complex parameters,
the $x^7,x^8,x^9$ 
coordinates  together with the gauge field component  $A_6$
in the $x^6$ coordinate. 
Let us count the dimensions of the baryonic
and non-baryonic 
branches in this type IIA picture.

\begin{figure}[htb]
\begin{center}
\includegraphics[width=3in]{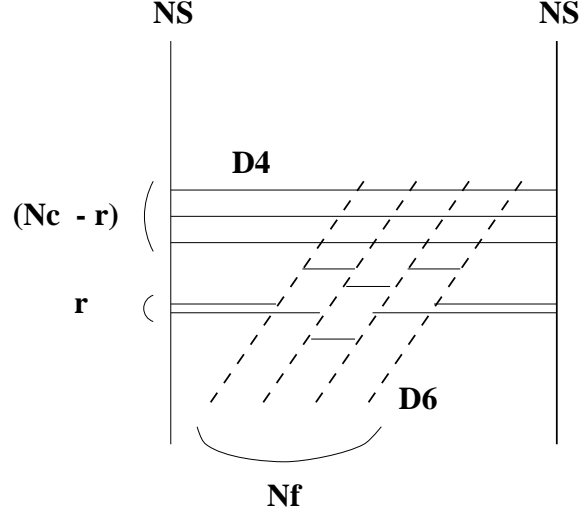}
\end{center}
\caption{The $r$-th non-baryonic branch in the type IIA picture.}
\label{fig2}
\end{figure}

The $r$-th non-baryonic branch, as depicted in figure \ref{fig2},
 corresponds to $(N_c - r)$  D4-branes suspended between
the two NS 5-branes and $r$ D4-branes
 broken on the D6-branes.

 Since the $s$-rule \cite{hw} 
does not allow more than one D4-brane
to be suspended between a NS 5-brane
 and a D6-brane, $r$ cannot be greater than $[N_f/2]$.
Since the Coulomb branch in the brane picture correponds to  
 \hbox{D4-branes} moving along the two NS 5-branes,
the $r$-th non-baryonic branch shares $(N_c-r-1)$ complex
dimensions with the Coulomb branch,
corresponding to gauge group $SU(N_c-r)$.
 The complex dimension of the non-baryonic
 branch in the Higgs direction is determined
by counting the number of D4-branes suspended 
between  the D6-branes. Taking into account the $s$-rule,
we obtain the dimension to be 
 \beq
 2 \big( (N_f-1) + (N_f-3) + ...+ (N_f-2r+1)  \big) = 2r(N_f-r)
 \comma
 \eeq
in agreement with the field theory results.
 
 \begin{figure}[htb]
\begin{center}
\includegraphics[width=4.2in]{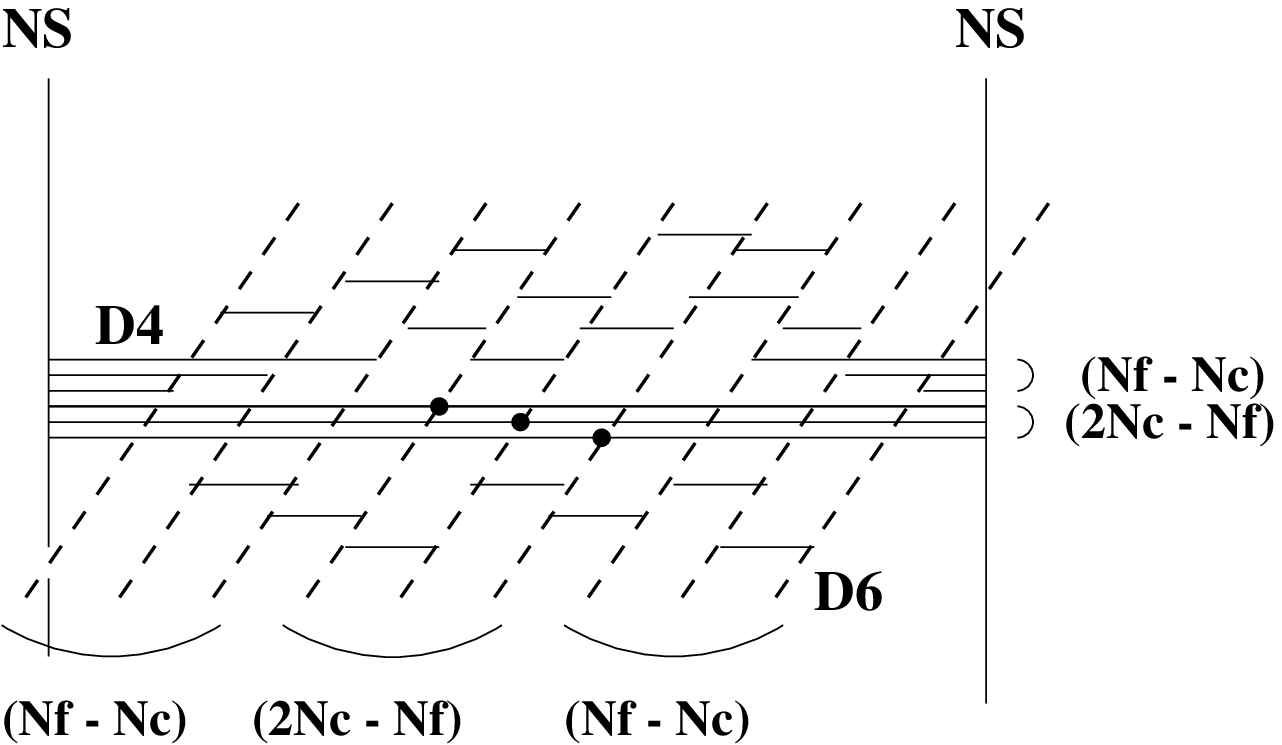}
\end{center}
\caption{The baryonic branch in the type IIA picture.}
\label{fig3}
\end{figure}

 The baryonic branch corresponds to complete Higgsing 
 as in figure \ref{fig3}.
 In this case counting the number of D4-brane pieces 
between two D6-branes 
 yields a complex dimension $2N_fN_c -2N_c^2$
for the baryonic branch. This is {\it not} 
the correct dimension. Compared to the field theory
result, we are missing 2 complex moduli.
We will show how the M-theory fivebrane description
accounts for these missing moduli.
 
The brane configuration can be 
reinterpreted in M-theory as a configuration of a single fivebrane 
with worldvolume $R^4 \times \Sigma$ where $\Sigma$ is some 
genus $(N_c-1)$ curve (Riemann surface). It was shown in \cite{witten}
that $\Sigma$ is nothing but the Seiberg-Witten curve \cite{sw}
that determines the structure of the Coulomb branch of the $N=2$
theory. The information on the meromorphic one-form (the Seiberg-Witten
form) on $\Sigma$ is carried by the embedding of $\Sigma$
in the space-time.

We follow the notation of \cite{witten} and set
$v=x^4+i x^5$, $s = (x^6 + i x^{10})/R$, $t = {\rm exp}(-s)$ 
where $x^{10}$ is the eleventh coordinate of M-theory
which is compactified on a circle
of radius $R$. The curve $\Sigma$ is given by an algebraic equation
in $(v,t)$ space. Specifically, for
$N=2$ $SU(N_c)$ gauge theory with $N_f$ flavors, it is given by
\beq
t^2 - C_{N_c}(v,u_k)t  + \Lambda_{N=2}^{2N_c-N_f}
  \prod_{i=1}^{N_f} (v +m_i) = 0
\comma
\label{cu}
\eeq 
where $C_{N_c}$ is a degree $N_c$ polynomial in $v$,
$C_{N_c} = v^{N_c} + \cdots$, with coefficients that depends on the 
moduli
$u_k$, and $m_i$ $(i=1,...,N_f)$ are the quark masses.

\subsection{The D6-Branes}

To describe the Higgs branch of the theory, 
it is useful to introduce the D6-branes in the system,
as we have seen in the Type IIA set-up.
In M-theory, 
the D6-branes are Kaluza-Klein Monopoles described by
a Taub-NUT space \cite{tow}.
One of the complex structures of this Taub-NUT space
is the same as one of the complex structures
of the ALE space of $A_{n-1}$-type:
\beq
yx = \Lambda_{N=2}^{2N_c-N_f}\prod_{i=1}^{N_f} (v+m_i) .
\label{tz}
\eeq
The D6-branes are located at $x=y=0$, $v=-m_i$.
In this framework, the Riemann surface $\Sigma$ is defined as a curve
in this surface given by
\beq
y+x=C_{N_c}(v,u_k)
\stop
\label{cc}
\eeq
It is easy to see that this description is the same as (\ref{cu})
under the identification $y=t$.

The Type IIA brane configuration is invariant under the rotations
in the $x^4,x^5$ and $x^7,x^8,x^9$ directions, which we denote
$U(1)_{4,5}$ and $SU(2)_{7,8,9}$ if the order parameters are also
rotated appropriately.
These are interpreted as the classical $U(1)$ and $SU(2)$
$R$-symmetry groups of the four-dimensional theory
on the brane worldvolume.
In the M-theory configuration, $SU(2)_{7,8,9}$
is preserved but $U(1)_{4,5}$ is broken.
We can preserve the discrete subgroup
$\Z_{4N_c-2N_f}$ \footnote{
Since the $\Z_2$ subgroup acts trivially on the space-time
coordinates and on the order parameters, we often
call this a discrete $\Z_{2N_c-N_f}$ R-symmetry.}
of $U(1)_{4,5}$ if we 
modify the $U(1)_{4,5}$ action so that the
variables $x$ and $y$ have charge $2N_c$.
The full $U(1)$ symmetry
is restored if we assign the instanton charge
$(4N_c-2N_f)$ to the factor $\Lambda_{N=2}^{2N_c-N_f}$,
reflecting the axial anomaly of the $U(1)_R$.
We list here the modified $U(1)_{4,5}$ charges
\beq
\begin{array}{cccc}
x&y&v&\Lambda_{N=2}^{2N_c-N_f}\\
2N_c&2N_c&2&4N_c-2N_f.
\end{array}
\label{45}
\eeq

When some $m_i$ coincide, corresponding D6-branes are located
at the same position in the $x^4,x^5$ directions, but they can be
separated in the $x^6$ direction \cite{hw}.
When $n$ of the bare masses are the same, the surface
(\ref{tz}) develops an $A_{n-1}$ singularity.
The separation of the D6-branes in the $x^6$ direction
corresponds to the resolution of this singularity \cite{witten}.
This resolution makes it possible to identify the Higgs branch
of the $N=2$ theory on the fivebrane worldvolume.
We now digress to give a brief description
of the resolution of the $A_{n-1}$ singularity
(see \cite{hov} for more detailed discussion).

\bigskip
{\it Resolution of $A_{n-1}$ Singularity}

The complex surface
\beq
yx=v^n
\eeq
embedded in the $x$-$y$-$v$ space has a singularity at the origin.
The resolution of this means a smooth complex surface
that is mapped onto this singular surface in such a way that
the map is an isomorphism except on the inverse image of the
singular point.
This is explicitly given as follows.

The resolved surface is
covered by $n$ complex planes $U_1$, $U_2$, $U_3$, ..., $U_n$
with coordinates $(y_1=y,x_1)$,
$(y_2,x_2)$, ..., $(y_n,x_n=y)$ which are
mapped to the singular $A_{n-1}$ surface by
\beq
U_i\ni(y_i,x_i)\longmapsto\left\{
\begin{array}{l}
y=y_i^ix_i^{i-1}\\[0.2cm]
x=y_i^{n-i}x_i^{n+1-i}\\[0.2cm]
v=y_ix_i ~.
\end{array}
\right.
\label{projA}
\eeq
The planes $U_i$ are glued together by $x_{i}y_{i+1}=1$ and
$y_{i}x_{i}=y_{i+1}x_{i+1}$.
The map onto the singular $A_{n-1}$ surface is an isomorphism except
on the inverse image of the singular point $x=y=v=0$.
The inverse image consists of $n-1$ $\CP^1$s $C_1$, $C_2$, ...,
$C_{n-1}$ where $C_i$ is the locus of $y_{i}=0$ in $U_{i}$
and $x_{i+1}=0$ in $U_{i+1}$, and is coordinatized by $x_i$ and
$y_{i+1}$ that are related by $x_iy_{i+1}=1$.
$C_i$ and $C_j$ do not intersect unless $j=i\pm 1$,
and $C_{i-1}$ and $C_i$ intersect transversely at $y_i=x_i=0$.
\begin{figure}[htb]
\begin{center}
\includegraphics[width=4.5in]{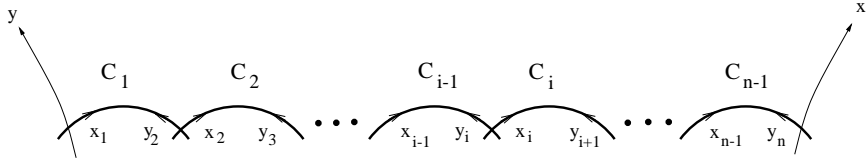}
\end{center}
\caption{Resolution of $A_{n-1}$ singularity.}
\label{figA}
\end{figure}

\medskip
In what follows in this section, we turn off the bare mass:
$m_i=0$ for any $i$. We separate the D6-branes in the $x^6$ direction
and
the eleven-dimensional space-time is then given by $\R^7$
times the resolved $A_{N_f-1}$ surface.
In this case,
the fivebrane is still described by the
same equation (\ref{cc}) where $x,y,v$ are now considered
as the functions (\ref{projA}) with $n=N_f$.
The position of the D6-branes are interpreted as
the $N_f$ intersection points of the rational curves
$C_1,C_2,\ldots,C_{N_f-1}$.

\subsection{The Higgs Branch}

In the Type IIA picture,
the Higgs branch is described by D4-branes suspended between 
D6-branes
where they can move in the $x^7,x^8,x^9$ directions.
Likewise,
in the present context, the transition to the Higgs branch occurs
when the fivebrane intersects with the D6-branes.
This is possible only
when the image (\ref{cc}) in the $x$-$y$-$v$ space of the curve
passes through the singular point $x=y=v=0$.
Thus, we must have $C_{N_c}(v=0)=0$, or in other words
$C_{N_c}(v)$ factorizes as
\beq
C_{N_c}(v)=v^r(v^{N_c-r}+u_2 v^{N_c-r-2}+\cdots+u_{N_c-r})\,,
\label{nbc}
\eeq
where $r>0$.

Now, we describe what this curve looks like in the resolved
$A_{N_f-1}$ surface. We first consider the case in which
the coefficients $u_2,\ldots,u_{N_c-r}$ are generic.
In particular $u_{N_c-r}\ne 0$.

It is convenient to look at the curve
by separating it into two pieces; one is near, the other is away from
the singularity $x=y=v=0$.
Away from the singular point
$x=y=v=0$, we can consider the curve
as embedded in the original $x$-$y$-$v$ space
because there is no distinction from the resolved surface in this
region.
As can be seen by
looking
at the two equations (\ref{tz})-(\ref{cc}), $v$ never
vanishes in this region of the curve.
Thus, we can safely divide the coordinates $x$ and $y$ by some
power of $v$. If $2r\leq n$ they can be divided by $v^r$,
while they can be divided by $v^{[N_f/2]}$ if $2r>N_f$. 
Then, we see that this piece of the curve is equivalent with the
generic curve
for the $SU(N_c-r)$ gauge theory with $(N_f-2r)$ flavors
and thus has genus $(N_c-r-1)$ for $2r\leq N_f$,
while it is some special genus $(N_c-[N_f/2]-1)$ curve
of the $SU(N_c-[N_f/2])$ gauge theory with $(N_f-2[N_f/2])$ 
flavors if $2r>N_f$.

\begin{figure}[htb]
\begin{center}
\includegraphics[width=5in]{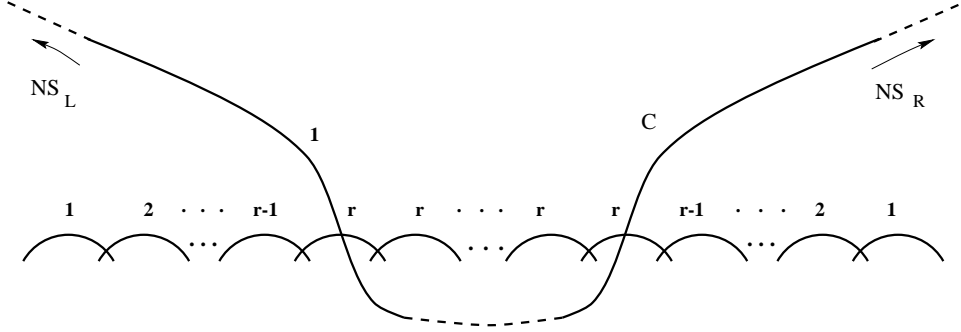}
\end{center}
\caption{Non-baryonic branch root in M-Theory.}
\label{fignbb}
\end{figure}

Near $x=y=v=0$, however, we must recall that we are actually
considering the resolved $A_{N_f-1}$ surface.
Thus, we must describe the curve in the $N_f$ patches as described
above. Before doing this, it is useful to remark that the higher
order terms $v^{r+1},v^{r+2},\ldots$ are negligible near $v=0$
compared to $v^r$. Thus, nothing essential is lost
if we replace the defining equation
$y+x=v^r(u_{N_c-r}+\cdots)=0$ by $y+x=v^r$.
On the $i$-th patch $U_i$, the equation looks like
\beq
y_i^ix_i^{i-1}+y_i^{N_f-i}x_i^{N_f+1-i}=y_i^rx_i^r\,.
\eeq
If $N_f\geq 2r$, one of the three terms is of lowest order
both in $y_i$ and $x_i$:
For $1\leq i\leq r$, the lowest order term is the first term on the
LHS,
for $r+1\leq i\leq N_f-r$, it is the term on the RHS, and
for $N_f-r+1\leq i\leq N_f$ it is the second term on the LHS.
Thus, the equation factorizes as
\begin{equation}
\begin{array}{lr}
y_i^ix_i^{i-1}(1+y_i^{N_f-2i}x_i^{N_f-2i+2}-y_i^{r-i}x_i^{r+1-i})=0 &\\
i=1,\ldots,r
& \\[0.2cm]
y_i^rx_i^r(y_i^{i-r}x_i^{i-r-1}+y_i^{N_f-r-i}x_i^{N_f+1-r-i}
-1)=0 &\\
i=r+1,\ldots,N_f-r
& \\[0.2cm]
y_i^{N_f-i}x_i^{N_f+1-i}
(y_i^{2i-N_f}x_i^{2i-N_f-2}+1-y_i^{i-N_f+r}x_i^{i-N_f+r-1})=0 &\\

i=N_f-r+1,\ldots,N_f ~~ .
&\\[0.2cm]
\end{array}
\end{equation}
We see that the curve consists of several components.
One component, which we call $C$, is the zero of the last factor of
the above equations. This extends to the one in
the region away from $x=y=v=0$ which we have already considered.
The other components are the rational curves $C_1,\ldots,C_{N_f-1}$.
Recall that $C_i$ is defined by $y_i=0$ in $U_i$ and $x_{i+1}=0$
in $U_{i+1}$.
In general, these have multiplicities. As is evident from the
above factorized form, the component $C_i$ has multiplicity $\ell_i$
where
$\ell_i=i$ for $i=1,\ldots,r$, $\ell_i=r$ for $i=r+1,\ldots,N_f-r$, 
and
$\ell_i=N_f-i$ for $N_f-r+1,\ldots,N_f-1$. Note that the component 
$C$
intersects with $C_r$ and $C_{N_f-r}$ as we can see by looking at the
equation at $i=r,r+1$ and $i=N_f-r,N_f-r+1$.
The curve is depicted in  figure \ref{fignbb}.

If $2r>N_f$, the structure of degeneration of the curve is the same
as the case $r=[N_f/2]$.
Recalling the behavior away from $x=y=v=0$,
we conclude that the cases $2r>N_f$ can be considered
as some special cases of $r=[N_f/2]$.

As noted in \cite{witten},
once the curve degenerates and $\CP^1$ components are generated,
they can move in the $x^7,x^8,x^9$ directions.
This motion together with the integration of the chiral two-forms
on such $\CP^1$'s parametrize the Higgs branch of
the four-dimensional theory.
Since the $\CP^1$ components are $\ell_i C_i$,
$i=1,\ldots,N_f-1$,
the quaternionic dimension of the $r$-th Higgs branch is
\beq
\sum_{i=1}^{N_f-1}\ell_i=2\sum_{i=1}^{r-1}
i+(N_f-2r+1)\times r=r(N_f-r).
\eeq
In view of the fact that there are $(N_c-r-1)$ parameters to deform the
infinite component $C$ in the $x$-$y$-$v$ direction, we can identify
this as the $r$-th non-baryonic branch emanating from an
$(N_c-r-1)$-dimensional subvariety of the Coulomb branch, as
introduced in \cite{aps}.

If we look at figure \ref{fignbb},
it is evident how
to identify the corresponding configuration in the
Type IIA picture. The $\CP^1$ components $\ell_i C_i$
correspond to the $\ell_i$
D4-branes stretched between the $i$-th and $i+1$-th D6-branes.
That these are the only allowed configurations of D4-branes 
gives a {\it proof} of the $s$-rule, conjectured in  \cite{hw},
which forbids more than one D4-brane to be stretched between
a NS 5-brane and a D6-brane. Further discussion on the $s$-rule
will be given in sections 5 and 6.

\bigskip
{\it The Baryonic Branch}

For the case $N_f\geq N_c$,
in addition to the non-baryonic branches, there is a baryonic branch
in which the gauge group is completely Higgsed \cite{aps}.
It has quaternionic dimension $N_fN_c-(N_c^2-1)$ and it 
emanates from a point in the Coulomb branch.
Here we look at the curve at the baryonic branch root, and
see how the transition to the baryonic branch is possible.

One of the basic property of the baryonic branch root
is that it is invariant under the discrete $R$-symmetry group
$\Z_{2N_c-N_f}$.
This requires all the color Casimirs $u_k$ to be
vanishing\footnote{According to the convention of
\cite{aps} where $C_{N_c}(v)=\prod_{a=1}^{N_c}(v-\phi_a)$,
the root of the baryonic branch is at
$\phi = (0, ..., 0, \omega, ..., \omega^{2N_c-N_f})$
where $\omega = e^{2\pi i/(2N_c-N_f)}$.}.
In this case
$C_{N_c}(v)=v^{N_c}+v^{N_f-N_c}$.
Thus, it is one of the non-baryonic branch roots with $r_*=N_f-N_c$
(note that $N_f-2r_*=2N_c-N_f\geq 0$).
The equation $y+x=v^{N_c}+v^{N_f-N_c}$ then
factorizes as
\beq
(y-v^{N_c})(y-v^{N_f-N_c})/y=0\,.
\eeq
Namely, the infinite curve $C$ factorizes into two rational curves
--- $C_L$ and
$C_R$ corresponding to $y=v^{N_c}$ and $y=v^{N_f-N_c}$
respectively.
\begin{figure}[htb]
\begin{center}
\includegraphics[width=5in]{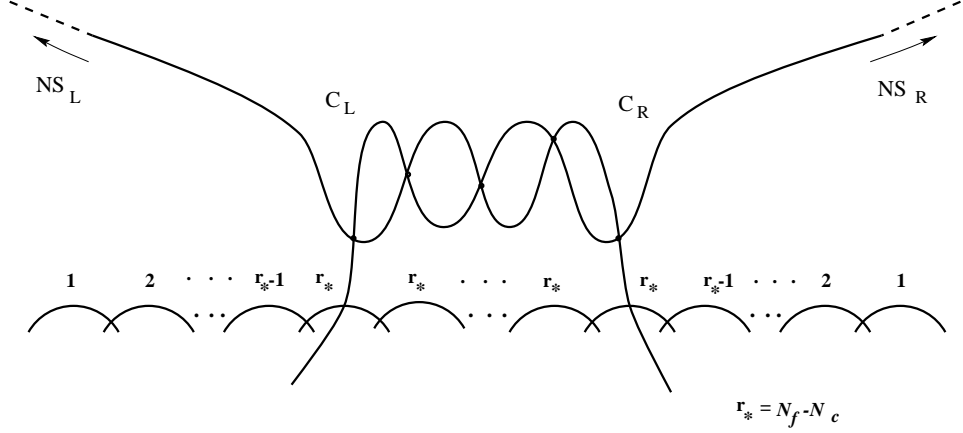}
\end{center}
\caption{The baryonic branch root in M-Theory.}
\label{figbb}
\end{figure}

The dimension of the baryonic branch is greater than that
of the \hbox{$r_*$-th} non-baryonic branch by 1 (in quaternionic
dimension). The former is $N_fN_c-(N_c^2-1)$ while the latter
is $r_*(N_f-r_*)=N_fN_c-N_c^2$. This difference corresponds
to the missing dimension of the baryonic branch in
the type IIA picture. How can we account for this difference?
We note that the two curves $C_L$ and $C_R$ intersects
at $(2N_c-N_f)$ points. This follows from the defining
equations $y=v^{N_c}$ and $y=v^{N_f-N_c}$
and the fact that they never intersect near $x=y=v=0$.
On the other hand, the infinite curve $C$ at the generic point
of the $r_*$-th
non-baryonic branch root has genus $N_c-r_*-1=2N_c-N_f-1$.
This means that, as the curve approaches 
the baryonic branch root,
it degenerates at $(2N_c-N_f)$ points and factorizes
into two rational curves. Thus
the curve $C$ at the baryonic branch root (or equivalently
the union of $C_L$ and $C_R$) describes the abelian gauge theory
with gauge group $U(1)^{2N_c-N_f-1}=\prod_i^{2N_c-N_f-1} U(1)_i$.
There are $(2N_c-N_f)$ massless electrons with charges
$(-1,0,\ldots,0)$, $(1,-1,0,\ldots,0)$, $\ldots,$
$(0,\ldots,1,-1)$, and $(0,\ldots,0,1)$ coming from the
degeneration points on the curve (the charges can be
read off from the intersection relations of the vanishing cycles). 
It is easy to see that such 
a theory has one-dimensional Higgs branch.
In this way, we have identified the missing $+1$ of the dimension
from the M-theory point of view.

\section{Rotating the Brane Configuration}

\newcommand{\tilC}{\widetilde{C}}
By adding a mass term to the adjoint chiral multiplet
in the $N=2$ vector multiplet, the $N=2$ supersymmetry
is broken to $N=1$.
In this section, we study the corresponding configuration
of fivebranes in M-theory.
In the Type IIA picture, 
this corresponds to changing the relative orientation
of the NS 5-branes.

\begin{figure}[htb]
\begin{center}
\includegraphics[width=3.5in]{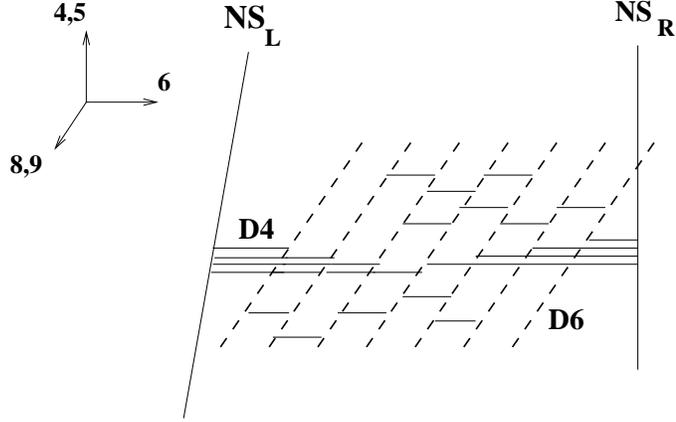}
\end{center}
\caption{Rotating the left NS 5-brane.}
\end{figure}

Since one of the 5-branes is going to be
extended in the
$(x^8, x^9)$ directions also, there is
only $N=1$ supersymmetry left on ${\mathbb{R}}^{1,3}$. 
To describe the corresponding
configuration of the M-theory fivebrane, let us introduce
a complex coordinate
\beq
    w = x^8 + i x^9.
\eeq
Before breaking the $N=2$ supersymmetry, the fivebrane is 
located at $w=0$.
We then rotate the left NS 5-brane toward
the $w$ direction while keeping the right NS 5-brane intact.
Since the two NS 5-branes correspond to the two asymptotic regions
with $v\to \infty$ where
$t=y\sim v^{N_c}$ and $x\sim v^{N_c}$ respectively where
the latter is equivalent to
$t\sim \Lambda_{N=2}^{2N_c-N_f}v^{N_f-N_c}$,
the rotation means that we impose the boundary condition as
\beq
\begin{array}{ccl}
w & \rightarrow & \mu v~~~~~{\rm as}~
v \rightarrow \infty,~
t \sim v^{N_c}\\[0.2cm]
w & \rightarrow & 0~~~~~~{\rm as}~
v \rightarrow \infty,~
t \sim \Lambda_{N=2}^{2N_c-N_f}v^{N_f-N_c} ~.
\end{array}
\label{cond}
\eeq

We can identify $\mu$ as the mass of the adjoint chiral multiplet
by using the $R$-symmetries.
Recall that the $N=2$ configuration is invariant under
the rotation groups
$U(1)_{4,5}$ and $SU(2)_{7,8,9}$ corresponding to the
$R$-symmetry of the field theory on the brane worldvolume,
where the action of $U(1)_{4,5}$ is modified as (\ref{45}).
After the rotation, $SU(2)_{7,8,9}$
is broken to $U(1)_{8,9}$ if the parameter $\mu$ in (\ref{cond})
is assigned the $U(1)_{4,5}\times U(1)_{8,9}$
charge $(-2,2)$.
Since this is the same as the $R$-charges of the mass of
the adjoint field and since there is no other parameter
charged with respect to $U(1)_{8,9}$, the two quantities 
should be identified.
We list below the charges of the coordinates and parameters.
\beq
\begin{array}{cccl}
&U(1)_{4,5}&U(1)_{8,9}&\\
v&2&0&\\
w&0&2&\\
y=t&2N_c&0&\\
x&2N_c&0&\\
\mu&-2&2\\
\Lambda_{N=2}^{2N_c-N_f}&4N_c-2N_f&0& ~.
\end{array}
\label{list}
\eeq

As a preliminary remark,
recall that in general the curve consists of several
components at the Higgs branch root. To rotate the curve 
in such a case, we pick the component, called $C$, that extends to 
infinity in the $v$ direction and rotate that component.
If the curve is not at the Higgs branch root,
the curve to be rotated is $\Sigma$ itself, but we denote
it as $C$ also in this case. 
At the baryonic branch root, the component $C$ further
factorizes into two rational curves $C_L$ and $C_R$.
In this case, the rotation is actually much easier than the other
cases, but we shall give a separate discussion
in view of its importance in a later section.

We should not expect to be able to rotate the brane configuration
(\ref{cu}) for arbitrary values of $u_k$'s. This is obvious
from the field theory point of view since the adjoint
mass lifts the Coulomb branch of the $N=2$ theory and the 
$u_k$'s are drawn to roots of Higgs branches. It is also clear
from the classical brane picture. If the D4-branes suspended
between the NS 5-branes are apart, the
two NS 5-branes have to remain parallel (if we force
the NS 5-branes to change their relative orientations,
the D4-branes get twisted and the supersymmetry is
completely broken).
In the M-theory picture, it is possible to rotate
a brane only when all the handles of the curve $C$ degenerate 
and $C$ can be considered as a single cylinder
with some points pairwise attached.
This can be seen by the following argument.

Suppose a curve $C$ is rotatable.
First, we note that the projection of
the rotated curve $\tilC$
on the $t$-$v$ plane remains the same
as (\ref{cu}).
This follows from the conservation of the $U(1)_{8,9}$ symmetry.
When $\mu$ is small, $w$ on $\tilC$ can be expressed 
as a function of $v$ and $t$. Thus the projection
of $\tilC$ on the $t$-$v$ plane can still be described by some
equation in $(t,v)$. To see that this equation is
the same as (\ref{cu}), we note that
$\mu$ is the only parameter that carries the $U(1)_{8,9}$
charge. Therefore we cannot deform the equation 
(\ref{cu}) without breaking the $U(1)_{8,9}$ symmetry. 
This means that the rotated curve $\tilC$ can be considered as the graph
of the ``function'' $w$ of the original curve $C$.
In order to clarify the property of this function $w$, we compactify
the curve $C$.
We note that both of the two asymptotic regions
--- $v\to \infty, t\sim v^{N_c}$ and
$v\to \infty, t\sim v^{N_f-N_c}$(\footnote{
We put $\Lambda_{N=2}=1$ for a while for simplicity.}) ---
are well parametrized by
the coordinate $v$.
Thus, we can compactify the curve by attaching the two
points {\it at infinity} $1/v=0$,
obtaining a compact Riemann surface $\overline{C}$.
The asymptotic condition (\ref{cond}) is equivalent 
to saying 
that $w$ is a meromorphic function of $\overline{C}$
which has only a simple pole at one point (one of the points
at infinity). Such a function exists only when the curve
$\overline{C}$ is equivalent to $\CP^1$.

\begin{figure}[htb]
\begin{center}
\includegraphics[width=4.0in]{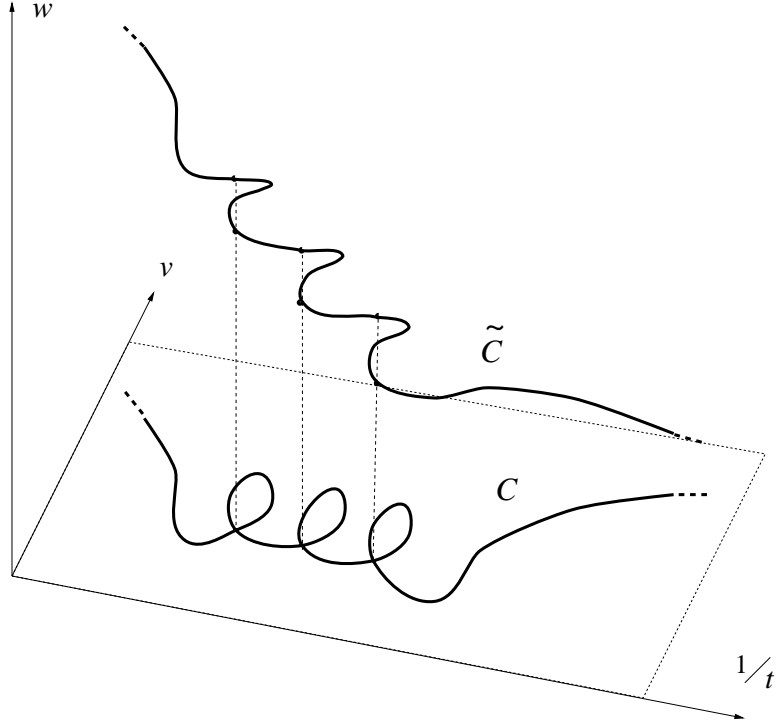}
\end{center}
\caption{Rotated curve as a graph of $w$.}
\label{figg}
\end{figure}

This completes the proof that the rotatable curve
$C$ should be completely degenerate (i.e. all the 
$\alpha$-cycles vanish). The proof
based on the asymptotic condition
(\ref{cond}) also shows that the rotated curve
$\tilC$ is a cylinder which is {\it globally} parametrized by
the coordinate $w$.
The two points at infinity of the curve correspond to
$w=0$ and $w=\infty$.

Thus, we can express $t$ and $v$ in terms of $w$ by rational functions:
\beq
v=P(w),~~~t=Q(w)\,.
\eeq
Since $v$ and $t$ never diverge except at the infinity $w=0,\infty$,
these rational functions are polynomials in $w$ up to
a factor of some power of $w$: $P(w)=w^a p(w)$, $Q(w)=w^b q(w)$
where $a$ and $b$ are some (possibly negative) integers
and $p(w)$ and $q(w)$ are polynomials of $w$
which we may assume non-vanishing at $w=0$.
Near one of the points at infinity $w=\infty$, $v$ and $t$ behave as
$v\sim \mu^{-1}w$ and $t\sim v^{N_c}$
by (\ref{cond}).
Thus, the rational functions are of the form
$P(w)=w^a(w^{1-a}+\cdots)/\mu$
and $Q(w)=\mu^{-N_c}w^b(w^{N_c-b}+\cdots)$. 
Let us look at the other infinity $w=0$ and consider
the cylinder to be compactified at this point.
A neighborhood of this point
is well parametrized by
$1/v$ which takes the value zero at $w=0$.
Recall that $w$ is the global coordinate of the
cylinder and hence
extends to the global coordinate of the compactified cylinder.
Namely, $1/v$ and $w$ are two good coordinates that vanish at the same
point. Thus they must be
linearly related
$w\sim {\rm const}/v$ in the limit $w\to 0$.
The function $P(w)$ then is of the form
\beq
P(w)=\frac{w^2+\cdots}{\mu w}=\frac{(w-w_+)(w-w_-)}{\mu w}\,.
\label{Pw}
\eeq
Since $t\sim v^{N_f-N_c}$ and $w\sim {\rm const}/v$ as $w\to 0$, 
we obtain $b=N_c-N_f$ and thus
$Q(w)=\mu^{-N_c}w^{N_c-N_f}(w^{N_f}+\cdots)$.
For $N_f>0$, by the equation $yx=v^{N_f}$ defining the space-time,
$t=0$ (i.e. $y=0$) implies $v=0$. Therefore the zeros of
the polynomial $w^{N_f}+\cdots$ coincide with the zeros $w_+$ and
$w_-$ of $P(w)$.
This way, we have determined the form of $Q(w)$ also:
\beq
Q(w)=\mu^{-N_c}w^{N_c-N_f}(w-w_+)^r(w-w_-)^{N_f-r}
\label{Qw}
\eeq
for some $r=0,1,\cdots, [N_f/2]$. We have imposed the bound $r\leq
[N_f/2]$ because the reflection $r\leftrightarrow N_f-r$ is
compensated by the exchange $w_+\leftrightarrow w_-$.

For $r>0$, both 
$v$ and $t$ get small near $w=w_{\pm}$
and behave as $t\sim v^r$ and $t\sim v^{N_f-r}$.
This is the property of the curve at the $r$-th baryonic branch root,
as can be seen by looking at the relation
(\ref{cu}) between $v$ and $t$ near $t=v=0$
which is approximately $t^2-v^rt+v^{N_f}=0$.
To see this in another way, we note that the component $C$
at the $r$-th non-baryonic branch root
intersects the exceptional
curve $x=y=v=0$ at two points near which $y=t$ and $v$ 
behave as $t\sim v^{r}$ and $t\sim v^{N_f-r}$.
Thus, we conclude that
the $r>0$ curve given by (\ref{Pw}) and (\ref{Qw}) 
is the rotation of the component $C$
at one of the $r$-th non-baryonic branch roots.
For $r=0$,
there is a point ($w=w_+$) at which $v=0$ but $t\ne 0$.
In this case the function $C_{N_c}(v)$ does not vanish at
$v=0$, which means that the curve does not pass through the
D6-branes in $x=y=v=0$. 
This is possible only if the curve is not at a Higgs branch
root.

The values $w_{\pm}$ are determined by the condition that
$v=P(w)$
and $t=Q(w)$ with (\ref{Pw})-(\ref{Qw})
satisfy identically the relation $t+v^{N_f}/t=C_{N_c}(v)=
v^rC_{N_c-r}(v)$.
This reads as
$$
w^{2N_c-N_f}(w-w_-)^{N_f-2r}
+\mu^{2N_c-N_f}(w-w_+)^{N_f-2r}=$$
$$=(w-w_+)^{N_c-r}(w-w_-)^{N_c-r}
+u_2(\mu w)^2(w-w_+)^{N_c-2-r}(w-w_-)^{N_c-2-r}+$$
$$+\cdots +u_{N_c-r}(\mu w)^{N_c-r}$$
up to the factor $(w-w_+)^r(w-w_-)^r/(\mu w)^{N_c}$.
The matching of the subleading and the lowest order terms
in powers of $w$ requires
(recovering $\Lambda_{N=2}$)
\beqa
&&\left(1-{r\over N_c}\right)w_++\left(1-{N_f-r\over N_c}\right)w_-
=0\,,
\label{lin}\\[0.2cm]
&&(w_+)^{N_c-N_f+r}(w_-)^{N_c-r}
=(-1)^{N_f}(\mu \Lambda_{N=2})^{2N_c-N_f}
\label{nolin}
\eeqa
and these determine $w_+$ and $w_-$ up to a $\Z_{2N_c-N_f}$
phase rotation. Matching of other terms determines
the coefficients $u_2,\ldots, u_{N_c-r}$, and these are uniquely
expressed in terms of $w_{\pm}$ so that the $\Z_{2N_c-N_f}$ action
on $w_{\pm}$
leads to the natural action on $u_i$.
Note that in the case $r=N_f/2$ ($N_f$ even),
the action of the $\Z_2$ subgroup is identified with the exchange of
$w_{+}$ and $w_-$ which does nothing on the rotated curve.

\noindent
{\bf Remark on the $r=r_*$ case:}
If $N_f\geq N_c$, there is a subtlety concerning the case
in which $r$ is $r_*=N_f-N_c$. For this value of $r$,
the equation (\ref{lin}) implies $w_+=0$. That is, the curve
is given by $v=(w-w_-)/\mu$, $t=(w-w_-)^{N_c}/\mu^{N_c}$
and one of the asymptotic regions
$v\to \infty$, $t\sim v^{N_f-N_c}$ is absent.
This means that such a curve cannot be realized as a curve of the
form (\ref{cu}). Thus, we must exclude this case from the list of
rotated curves. Actually, this is one of the components
of the rotated curve at the baryonic branch which we now
describe.

\medskip

{\it The Baryonic Branch}

The rotation of the curve at the baryonic branch root is straightforward.
The component $C$ at the baryonic branch
factorizes into two pieces ---
$C_L$ described by $t=v^{N_c}$, $w=0$ and $C_R$ described by
$t=v^{N_f-N_c}$, $w=0$.
The rotation
can be done just by
replacing $w=0$ for $C_L$ by
$w=\mu v$.
The curve is explicitly given by
\beq
\widetilde{C}_L\,
\left\{
\begin{array}{l}
t=v^{N_c}\\
w=\mu v
\end{array}
\right.
\qquad
C_R\,
\left\{
\begin{array}{l}
t=\Lambda_{N=2}^{2N_c-N_f}v^{N_f-N_c}\\
w=0 .
\end{array}
\right.
\label{rotb}
\eeq

\bigskip

\noindent
{\bf Summary}

\medskip
To summarize, we have identified all possible curves that can be
rotated.
The curve at the baryonic branch can be rotated, and the result is
given by (\ref{rotb}).
Some curves at the non-baryonic branch roots and some curves away
from the Higgs branch roots are also rotatable. 
The result of the rotation of these curves 
is given by the equations
($r=0,1,\ldots,[N_f/2]$, $r\ne r_*$)
\beqa
v&=&\frac{(w-w_+)(w-w_-)}{\mu w}\,,
\label{vw}\\[0.2cm]
t&=&\mu^{-N_c}w^{N_c-N_f}(w-w_+)^r(w-w_-)^{N_f-r}\,,
\label{tw}
\eeqa
where $w_{\pm}$ are determined by (\ref{lin}) and (\ref{nolin}).
For each $r<N_f/2$, $r\ne r_*$, there are $(2N_c-N_f)$ solutions related
by the $\Z_{2N_c-N_f}$ action, while for
$r=N_f/2$ ($N_f$ even) there are $(N_c-N_f/2)$ solutions.
Before the rotation, the curve for $r>0, r\ne r_*$
is at the $r$-th non-baryonic
branch root, while the curve for $r=0$ is not at a Higgs branch root.

We can perform a similar analysis in  the case in which the
quark mass term $m_f\sum_i \tilQ^iQ_i$ is turned on.
For generic values of $m_f$,
there is no baryonic branch and the curve $C$ never
factorizes.
We only have to repeat the same procedure
for the unfactorized curves by replacing $v$ by
$v+m_f$, and
the result is a slight modification of (\ref{vw})-(\ref{tw}) with
(\ref{lin})-(\ref{nolin}). That is, we replace (\ref{vw}) and
(\ref{lin}) by
\beqa
&&v+m_f=\frac{(w-w_+)(w-w_-)}{\mu w}\,,
\label{vwm}\\[0.2cm]
&&\left(1-{r\over N_c}\right)w_++\left(1-{N_f-r\over N_c}\right)w_-
+\mu m_f=0\, .
\label{linm}
\eeqa
For every $r=0,1,\ldots,[N_f/2]$, there are $2N_c-N_f$ solutions%
\footnote{
For special values of $m_f$, there are $r$ such that
$w_+$ or $w_-$ vanishes.
These presumably correspond to the vacua with meson matrix having
Jordan blocks which is mentioned in a footnote in section 2.}.

\bigskip

\noindent
{\bf Comparison with Field Theory}

The fivebrane configuration we found here encodes various 
information on the moduli space of vacua of the $N=1$ gauge theory.
Let us compare it with the results of the field theory
analysis in section 2.  
Here we will restrict our 
attention to the regions of the moduli space where the
vev of the baryon fields $B,\tilde{B}$ vanish,
and postpone the
discussion of non-zero vev
for $\tilde{B}B$ to section 6. 
We will find that the brane 
provides us with an exact description of the moduli space
of vacua.

\medskip

\bigskip
{\it Interpretation of $w_{\pm}$}

\medskip
Let us discuss the meaning of the values of $w_{\pm}$.
In the Type IIA set-up in which all the D6-branes are sent to
the infinity $x^6=+\infty$,
there are $N_f$ semi-infinite D4-branes ending on the right NS 5-brane
from the right. The fact that $t=v=0$ at $w=w_{\pm}$
means that $w=w_{\pm}$
are the asymptotic positions in the $w=x^8+ix^9$ direction of these
semi-infinite D4-branes.
Moreover, the order of zero in (\ref{tw}) says that
$r$ of the $N_f$ D4-branes are at $w=w_+$
in the limit $x^6\to +\infty$ and the remaining
$N_f-r$ are at $w=w_-$.

A $U(N_f)$ symmetry is associated with these semi-infinite D4-branes.
From the point of view of the four-dimensional field theory
on the D4-branes which are finite in the $x^6$ direction,
this appears as the global symmetry.
When the D4-branes are separated from each other,
the global symmetry is broken.
If the separation is in the $v=x^4+ix^5$ direction,
it is interpreted as the explicit breaking due to the bare mass
$(m_f)_i$ of the quarks, since these
are the only parameters charged under $U(1)_{4,5}$ that can
break the $U(N_f)$ flavor symmetry.
If the separation is in the $w$ direction,
it must be interpreted as due to a quantity
with $U(1)_{8,9}$ charge 2 that can break the $U(N_f)$ flavor
symmetry.
The only such quantity is the meson vev $M^i_j=\tilQ^iQ_j$.
{\it Thus, we interpret the position in the $w$ direction
of the semi-infinite D4-branes as the eigenvalues of the meson matrix
$M^i_j$}.

This interpretation can be verified as follows.
First, the fact that there are at most two values $w_{\pm}$
for the asymptotic $w$ position of the $N_f$ D4-branes is 
consistent with the field theory result
that there are at most two eigenvalues $m_{(1)}$ and $m_{(2)}$ 
of the meson matrix.
Moreover, for each degeneracy type $r$,
the positions $w_+$
and $w_-$ of degeneracy $r$ and $N_f-r$
agree with the eigenvalues
$m_{(1)}$ and $m_{(2)}$ of degeneracy $r$ and $N_f-r$ respectively,
up to an overall phase that depends only on $N_c$. 

Let us compare these two quantities in more detail.

\bigskip
\noindent
\underline{$0<N_f<N_c$}

Consider first the massless case.
The extrema of the exact superpotential (\ref{WM})
were found to satisfy equations (\ref{la}) and
(\ref{laa}) and solved by (\ref{sol}). The solution
for $r=0$ in (\ref{r0}) is a special case
of (\ref{sol}).
The equations defining $w_{\pm}$  (\ref{lin}) and (\ref{nolin})
are identical to 
(\ref{la}) and (\ref{laa}) up to an overall phase factor; thus we have
\beq
w_+ = (-1)^{\frac{N_c}{2N_c-N_f}}m_{(1)},~~~~~~~~ 
w_- = (-1)^{\frac{N_c}{2N_c-N_f}}m_{(2)}
\stop
\label{wm}
\eeq

Given the interpretation of $w_{\pm}$ as the
eigenvalues of the meson matrix, we see that 
the fivebrane in M-theory describes correctly the moduli space
of vacua on the $N=1$ gauge theory which is obtained in field theory
as the extrema of (\ref{WM}), including the dynamical generation of the 
Affleck-Dine-Seiberg superpotential and the non-renormalization of (\ref{WM}).

Consider next the massive  case.
The extrema of the exact superpotential (\ref{WMm})  
satisfy equations (\ref{1st}) and (\ref{2nd}).
The equations defining $w_{\pm}$  
(\ref{linm}) and (\ref{nolin})
are, as in the massless case, identical to 
(\ref{1st}) and (\ref{2nd}) up to an overall phase factor
and $w_{\pm}$ are related to $m_{(1)},m_{(2)}$
by (\ref{wm}).
We see that also in the massive case 
the fivebrane in M-theory describes correctly the exact moduli space
of vacua on the $N=1$ gauge theory.

\bigskip
\noindent
\underline{$N_f=N_c$}

The extrema of the superpotential (\ref{sup}) 
were found to satisfy equations (\ref{la1}) and
(\ref{la2}) and solved by (\ref{r}). 
The equations defining $w_{\pm}$  (\ref{lin}) and (\ref{nolin})
are identical to 
(\ref{la1}) and (\ref{la2}) up to an overall phase factor and 
$w_{\pm}$ are related to $m_{(1)},m_{(2)}$
by (\ref{wm}).
In this case we do not have a field theory proof 
that the superpotential (\ref{sup}) is exact.
Since we expect the M-theory fivebrane to describe the exact moduli space
of vacua, this result may be regarded as an evidence for
the non-renormalization theorem of (\ref{sup}).
In the limit $\mu\rightarrow \infty$, the low energy
theory becomes $N=1$ SQCD with $N_f=N_c$. In this case the moduli space
of vacua is modified quantum mechanically and the singularity  
at the origin is resolved. Indeed, the brane captures this phenomenon
as we shall see below.

\bigskip
\noindent
\underline{$N_c+1 \leq N_f$}

In section 2 we used the magnetic description in order to find the vacua.

The extrema of the  superpotential (\ref{sup1}) 
satisfy equations (\ref{la}) and
(\ref{laa}), which are also
 the  equations defining $w_{\pm}$  (\ref{lin}) and (\ref{nolin}), again 
 up to an overall phase factor.
$w_{\pm}$ are related to $m_{(1)},m_{(2)}$
by (\ref{wm}) as in the previous cases.
In this case we also do not have a field theory proof that 
the superpotential (\ref{sup1}) is exact. Our result here using
the M-theory fivebrane suggests that it is.

To summarize: With  the interpretation of $w_{\pm}$ as the
eigenvalues of the meson matrix, what we showed is that
the fivebrane in M-theory describes the exact moduli space of vacua
and captures the quantum phenomena such
as   the dynamical generation of the 
Affleck-Dine-Seiberg superpotential for $N_f <N_c$
and the quantum modification
to the classical moduli space of vacua when $N_f=N_c$.
Moreover it provided us with non-renormalization theorems for superpotentials
which have not been proven by field theory methods.

\newcommand{\tilt}{\widetilde{t}}
\newcommand{\tily}{\widetilde{y}}
\section{The $\mu \rightarrow \infty$ Limit: $N=1$ SQCD}

In the last section, we have found the configuration of
the fivebrane for a finite value
of the adjoint mass $\mu$. In this section, we take the $\mu \rightarrow
\infty$ limit of this configuration and compare it with the known result
of the $N=1$ supersymmetric field theory. 

\begin{figure}[htb]
\begin{center}
\includegraphics[width=3.5in]{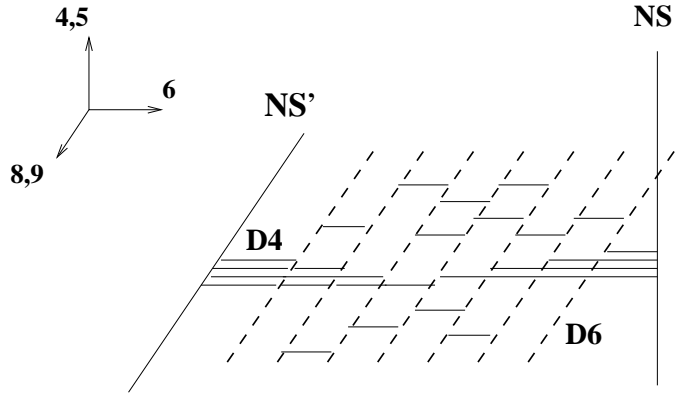}
\end{center}
\caption{The $\mu\to\infty$ limit of type IIA configuration.}
\end{figure}

Recall that we have been considering rotation of the left NS 5-brane
which corresponds to the asymptotic region
$t\sim v^{N_c}, v\to\infty$. After rotation, this region behaves as
$w\to \infty$, $v\sim\mu^{-1}w$, and
\beq
t\sim \mu^{-N_c}w^{N_c}\,.
\label{tww}
\eeq
Since the $N_c$ D4-branes end on the left NS 5-brane
also in the right angle limit,
we expect the relation $t\sim w^{N_c}$ to hold
in the limit $\mu\to \infty$ as well. 
For this, it is evident from (\ref{tww}) that we must rescale
$t$ by a factor $\mu^{N_c}$ and introduce the rescaled variable
\beq
\tilt=\mu^{N_c}t.
\label{tmu}
\eeq
Since the space-time is asymptotically a flat cylinder with flat
coordinate $x^6+ix^{10}=R\log (1/t)$, this rescaling 
simply corresponds to
the shift of the origin in the $x^6,x^{10}$ directions.
Moreover, the rescaling fits nicely with the symmetry property
and the renormalization group flow of the corresponding
four-dimensional physics, as we now see.

Recall that we are considering the eleven-dimensional
space-time to be a flat $\R^7$ times the Taub-NUT
space described by (\ref{tz}) where $y$ is identified with $t$. 
Using the rescaled variable (\ref{tmu}) or equivalently
$\tily=\mu^{N_c}y$, the
space-time is described by
\beq
\tily\, x=\mu^{N_c}\Lambda_{N=2}^{2N_c-N_f}
\prod_{i=1}^{N_f}(v+m_i)\,.
\label{N1ST}
\eeq
This expression has a smooth limit as $\mu\to \infty$
provided the constant $\Lambda_{N=1}$ given by
\beq
   \Lambda_{N=1}^{3N_c-N_f} = \mu^{N_c} \Lambda_{N=2}^{2N_c-N_f}
\label{matching2}
\eeq
is kept finite.
It may appear that this choice of limit
is ambiguous
because we could rescale the coordinate $x$ also.
However, this is not allowed since the fivebrane
behaves near the other
infinity $w\to 0$ as $x\sim v^{N_c}$, $v\to \infty$, and
this should also be preserved in the $\mu\to\infty$ limit.
The relation (\ref{matching2})
is the same as the renormalization group matching condition
of the corresponding four-dimensional field theory.
This space-time together with the fivebrane in it which we are going
to describe
is invariant under the rotation groups
$U(1)_{4,5}$ and $U(1)_{8,9}$ where
the charges
of the new parameters and coordinates are given by
\beq
\begin{array}{cccc}
&U(1)_{4,5}&U(1)_{8,9}&\\
v&2&0&\\
w&0&2&\\
\tily=\tilt&0&2N_c&\\
x&2N_c&0&\\[0.1cm]
\Lambda_{N=1}^{3N_c-N_f}&2N_c-2N_f&2N_c&\\[0.1cm]
\mu&-2&2&. 
\end{array}
\label{list1}
\eeq
In particular, the charges of the factor $\Lambda_{N=1}^{3N_c-N_f}$
mean that
the groups $U(1)_{4,5}$ and
$U(1)_{8,9}$
are broken to their discrete subgroups
$\Z_{2N_c-2N_f}$ and $\Z_{2N_c}$ respectively.
This
dictates precisely the anomaly of the corresponding
$U(1)$ $R$-symmetry groups.

\subsection{$SU(N_c)$ Without Matter}

In this case, the curve describing the
$N=2$ Coulomb branch is given by
\beq
 t^2 - C_{N_c}(v,u_k)t  + \Lambda_{N=2}^{2N_c} = 0.
\label{pure}
\eeq
There are $N_c$ points on the Coulomb branch where the curve
is completely degenerate, and these points are related to each
other by the unbroken discrete $\Z_{4N_c}$ subgroup
of $U(1)_{4,5}$ acting as
$v\to v \e^{{2\pi i\over 2N_c}}$, $t\to -t$
which has the effect
$\Lambda_{N=2}^2\to \Lambda_{N=2}^2 \e^{{2\pi i\over N_c}}$
on the curve.

At one of these points, the curve
takes the form,
\beq
  v = \Lambda_{N=2}^2 t^{-1/N_c} + t^{1/N_c},
\label{limit1}
\eeq
and its rotation is given by
\beqa
  v&=& \mu \Lambda_{N=2}^2 w^{-1} + \mu^{-1} w,
\nonumber \\
  t& = & \mu^{-N_c} w^{N_c}. 
\label{limit2}
\eeqa
Before taking the $\mu \rightarrow \infty$ limit, we rescale $t$ as
$\tilt=\mu^{N_c}t$ (\ref{tmu}).
The equations (\ref{limit1}) and (\ref{limit2})
can then be rewritten as
\beqa
  v &=& \Lambda_{N=1}^3 \tilt^{-1/N_c} +   
           \mu^{-1} \tilt^{1/N_c} ,\nonumber \\
  w & = & \tilt^{1/N_c},  \nonumber \\
  vw & = & \Lambda_{N=1}^3 + \mu^{-1} w^2 ,
\eeqa
where, following (\ref{matching2})
\beq
   \Lambda_{N=1}^3 = \mu \Lambda_{N=2}^2 .
\label{matching}
\eeq
This curve has a smooth $\mu \rightarrow \infty$ limit
if we send $\Lambda_{N=2} \rightarrow 0$ at the same time
so that $\Lambda_{N=1}$ remains finite. 
Dropping the terms multiplied by $\mu^{-1}$, the curve in this limit becomes
\beqa
  v &=& \Lambda_{N=1}^3 \tilt^{-1/N_c} , \nonumber \\
  w & = & \tilt^{1/N_c},\\
vw &=& \Lambda_{N=1}^3\,.
\label{purecurve}
\eeqa
We note that $\Lambda_{N=1}$ characterizes the size of
the fivebrane configuration (\ref{purecurve}).
The $\Z_{2N_c}$ subgroups of $U(1)_{4,5}$ and $U(1)_{8,9}$
have the same effect
\beq
\Lambda_{N=1}^3\to \e^{{2\pi i\over N_c}}\Lambda_{N=1}^3
\eeq
on the curve.
The $N_c$ curves, or fivebranes,
are related by this discrete $\Z_{2N_c}$
symmetry group which is spontaneously broken to $\Z_2$.
This is what we have observed in the four-dimensional
$N=1$ super-Yang-Mills theory.

\subsection{$SU(N_c)$ With $N_f$ Massless Matter}

\subsubsection{Non-Baryonic Branches}

In terms of the rescaled variables,
the equations describing the rotated brane
at the root of the $r$-th non-baryonic branch
($r=0,1,..,[N_f/N_c]$; \break
$r\ne N_f-N_c$) are
\beqa
&&\mu^{-N_c}\tilt^2-v^{r}C_{N_c-r}(v,u_k)\,\tilt
+\Lambda_{N=1}^{3N_c-N_f} v^{N_f}  = 0,
\label{limit4}\\[0.2cm]
&&\tilt=w^{N_c-N_f}(w-w_+)^r (w-w_-)^{N_f-r}, 
\label{limit5} \\[0.2cm]
&&vw = \mu^{-1} (w-w_+)(w-w_-) \ ,
\label{limit6} 
\eeqa
where $w_{\pm}$ are determined by (\ref{lin}) and (\ref{nolin}).
In particular, $w_{\pm}$ are of the form
\beq
w_{\pm}=c_{\pm}\mu\Lambda_{N=2} \ ,
\eeq
where $c_{\pm}$ are non-zero numerical constants depending on
$N_c,N_f$ and $r$. We recall that
$\Lambda_{N=2}$ and $\Lambda_{N=1}$ are related by
(\ref{matching2}) and we send $\Lambda_{N=2}$ to zero so that
$\Lambda_{N=1}$ is kept finite.
Since the order parameters $u_k$ are powers of $\Lambda_{N=2}$
and are independent of $\mu$, they vanish in the $\mu\to \infty$
limit, which means that $v^rC_{N_c-r}(v)\to v^{N_c}$.
Thus, the equation (\ref{limit4}) has the smooth limit
described by
\beq
     \tilt = \Lambda_{N=1}^{3N_c-N_f} v^{N_f-N_c} .
\label{limit7}
\eeq
How the other two equations (\ref{limit5}) and
(\ref{limit6}) behave in this limit depends on the behavior of
$w_{\pm}\sim\mu \Lambda_{N=2}$. Since
\beq
   \mu \Lambda_{N=2} = \left( \mu^{N_c-N_f} \Lambda_{N=1}^{3N_c-N_f}
   \right)^{\frac{1}{2N_c-N_f}}, 
\eeq 
we should distinguish three cases, $N_f<N_c$, $N_f=N_c$ and
$N_f>N_c$.

When $N_f<N_c$, $\mu \Lambda_{N=2}$ diverges in the limit $\mu
\rightarrow \infty$. This means that the curve defined by
(\ref{limit4}), (\ref{limit5}) and (\ref{limit6}) becomes infinitely
elongated along the $x^6$ direction. Thus the $\mu \rightarrow \infty$
limit of the brane configuration does not provide a field theory
in four dimensions. This is consistent with the field theory result
that there is no supersymmetric vacuum for the corresponding $N=1$
theory.

When $N_f> N_c$, $\mu \Lambda_{N=2}$ vanishes for $\mu \rightarrow
\infty$. In this limit,
the equations (\ref{limit4})-(\ref{limit6}) become
\beqa
     v^{N_c}\tilt &=& \Lambda_{N=1}^{3N_c-N_f} v^{N_f}~, \nonumber \\
     \tilt &=& w^{N_c}~, \nonumber
 \\ vw &=& 0 ~.
\eeqa
It may appear that only $\tilt=w^{N_c}$ and $v=0$ is allowed
since the second equation seems to imply that $w=0$ means $\tilt=0$.
However, the double limit $w\to 0$, $\mu\to\infty$
of the equation (\ref{limit5})
is subtle. As we will see later, $\tilt\ne 0$, $w=0$ is allowed.
Thus, the correct interpretation of these equations
is that the curve splits into two components
in this limit:
\beq
C_L\,\,
\left\{
\begin{array}{l}
\tilt= w^{N_c}\,,\\
v=0\,,
\end{array}
\right.
\qquad
C_R\,\,
\left\{
\begin{array}{l}
\tilt= \Lambda_{N=1}^{3N_c-N_f} v^{N_f-N_c}\,,\\
w=0\,.
\end{array}
\right.
\label{nf>nccurve}
\eeq
The component $C_L$ corresponds to the NS${}^{\prime}$
5-brane and the D4-branes attached to it, while
the component $C_R$ corresponds to the NS 5-brane and the attached
D4-branes.
It is interesting
to note that all the non-baryonic branch roots have the same limit.

To be more precise, there are other components corresponding to D4-branes
stretched between the D6-branes. These are rational curves
located at the exceptional divisor $x=y=v=0$.
In addition to the ones
which are present already before rotation (i.e. $N=2$ limit
$\mu=0$), there are also rational
curves which appear only in the $\mu\to
\infty$ limit. We will describe how they appear shortly.
It turns out that
all the non-baryonic branch roots
have the same limit as well even if we take these rational curves into
account.

When $N_f = N_c$, $\mu \Lambda_{N=1}$ is equal to $\Lambda_{N=1}^2$.
In the limit $\mu\to\infty$ the equations
(\ref{limit4})-(\ref{limit6})
become
\beqa
\tilt &=& \Lambda_{N=1}^{2N_c}~,  \nonumber \\
\tilt &=& (w-c_+ \Lambda_{N=1}^2)^r
(w-c_- \Lambda_{N=1}^2)^{N_c-r}~, \nonumber\\
vw &=& 0 ~ .
\eeqa
As in the case of $N_f>N_c$,
the correct interpretation of these is that
the curve splits into two components.
\beq
C_L\,\,
\left\{
\begin{array}{l}
\tilt=(w-c_+ \Lambda_{N=1}^2)^r
(w-c_- \Lambda_{N=1}^2)^{N_c-r}
\,,\\
v=0\,,
\end{array}
\right.
~
C_R\,\,
\left\{
\begin{array}{l}
\tilt= \Lambda_{N=1}^{2N_c}\,,\\
w=0\,.
\end{array}
\right.
\label{nf=nccurve}
\eeq
Unlike the case
of $N_f > N_c$, the final configuration of the curve depends
on $r$, i.e. different non-baryonic branch roots go to different
limits.
As in the $N_f>N_c$ case, there are other finite components.
As we will see, those which appear only in the $\mu\to\infty$
limit are located at $w=w_{\pm}$.

\subsubsection{Baryonic Branch}

The baryonic branch exists for $N_f \geq N_c$.
Before the right angle limit, the rotated curve is given by
(\ref{rotb}). In terms of the rescaled variables, it is written as
\beq
\widetilde{C}_L\,
\left\{
\begin{array}{l}
\tilt=w^{N_c}\\
v=\mu^{-1}w
\end{array}
\right.
\qquad
C_R\,
\left\{
\begin{array}{l}
\tilt=\Lambda_{N=1}^{3N_c-N_f}v^{N_f-N_c}\\
w=0 ~.
\end{array}
\right.
\label{bbmu}
\eeq
In the limit $\mu \rightarrow \infty$,
these become
\beq
C_L\,
\left\{
\begin{array}{l}
\tilt=w^{N_c}\\
v=0
\end{array}
\right.
\qquad
C_R\,
\left\{
\begin{array}{l}
\tilt=\Lambda_{N=1}^{3N_c-N_f}v^{N_f-N_c}\\
w=0 ~.
\end{array}
\right.
\label{bb}
\eeq
For $N_f > N_c$, the baryonic branch root has the same limit
as the non-baryonic roots given by (\ref{nf>nccurve}). Thus in 
this case, all the roots of the Higgs branches converge
to the same curve. On the other hand,
for $N_f=N_c$, (\ref{bb}) are different
from (\ref{nf=nccurve}) for the non-baryonic branches.
This reflects the fact that the quantum moduli space for the $N=1$
theory with $N_f=N_c$ is different from the classical one 
while the moduli space for $N_f>N_c$ does not receive
quantum corrections. This point will be discussed further in
the next section.

\subsection{$SU(N_c)$ With Massive Matter}

If $N_f<N_c$ and the quarks are massless, the $\mu \rightarrow \infty$
limit of the theory has no supersymmetric vacua
as we have just seen by brane analysis and also
in Section 2 from the point of view of field theory.

However we can
stabilize the vacuum by adding a quark mass term
which breaks $U(1)_{4,5}$ but preserves $U(1)_{8,9}$.
From the field theory
analysis, we have seen that the non-baryonic branch with only $r=0$
survives in this case. So let us examine how it looks like from
the point of view of the brane configuration. 

If all the quarks have the same mass $m_f$, the curve
for finite $\mu$ is
\beqa
   v + m_f & = & \frac{(w - w_+)(w - w_-)}{\mu w}~, \nonumber \\
   \mu^{N_c} t & = & w^{N_c-N_f} (w - w_-)^{N_f}~. 
\eeqa
In the limit $\mu \rightarrow \infty$, 
\beqa
   w_+ &\simeq&  - \mu m_f~, \nonumber \\
   w_- &\simeq&\left( 
\frac{\Lambda_{N=1}^{3N_c - N_f}}{m_f^{N_c-N_f}}\right)^{1/N_c}~. 
\eeqa
Therefore the curve in this limit becomes
\beqa
  v &=& \left( m_f^{N_f} \Lambda_{N=1}^{3N_c - N_f} \right)^{1/N_c}
  w^{-1}~,
  \nonumber \\
  \tilt&=& w^{N_c-N_f} \left(w -\left(
  \frac{\Lambda_{N=1}^{3N_c - N_f}}{m_f^{N_c-N_f}} \right)^{1/N_c}
\right)^{N_f} ~ ,
\label{curvewithmatter} 
\eeqa
where $\tilt = \mu^{N_c} t$ as before. The limit $\mu \rightarrow
\infty$ has been taken so that $\Lambda_{N=1}$ given in
(\ref{matching2}) remains finite.
Note that there are $N_c$ solutions related to each other
by the action of the discrete $\Z_{2N_c}$ subgroup of $U(1)_{8,9}$.

It is interesting to see how this configuration reduces to that
for $N_f=0$ in the limit $m_f \rightarrow \infty$.
If we define 
\beq
 m_f^{N_f} \Lambda_{N=1}^{3N_c - N_f} = \tilde{\Lambda}_{N=1}^{3N_c}
  ~,
\label{matching3}
\eeq
the curve (\ref{curvewithmatter}) can be written as
\beqa
  v &=&  \frac{\tilde{\Lambda}_{N=1}^3}{w}~,
  \nonumber \nonumber \\
  \tilt&=& w^{N_c-N_f} \left(w -
  \frac{\tilde{\Lambda}_{N=1}^3}{m_f} \right)^{N_f}~. 
\label{purecurve2} 
\eeqa
If we keep $\tilde{\Lambda}_{N=1}$ finite while we send $m_f \rightarrow
\infty$, the space-time in this limit is just the flat cylinder
given by
\beq
\tily\,x=\tilde{\Lambda}_{N=1}^{3N_c},
\eeq
and the fivebrane (\ref{purecurve2}) reduces
to the pure Yang-Mills result (\ref{purecurve}).
Once again,  (\ref{matching3}) is exactly the renormalization
matching condition for the corresponding situation in the field theory.      

On the other hand, the limit
$m_f \rightarrow 0$ of (\ref{curvewithmatter}) gives
an infinitely elongated curve (the branch at $w =
(\Lambda_{N=1}^{3N_c-N_f}/m_f^{N_c-N_f})^{1/N_c}$ goes
to the infinity). This corresponds to the fact that
there is no supersymmetric vacuum for the $SU(N_c)$ theory
with massless $N_f (< N_c)$ flavors.

\subsection{Generation of Rational Curves in the $\mu\to\infty$ Limit}

In the subsection 5.2.1, we mentioned that extra rational curves
appear in the $\mu \rightarrow \infty$ limit. We will show this in
this subsection. In the Type IIA picture, this corresponds
to breaking the D4-branes attached to the NS${}^{\prime}$ 5-brane
at the D6-branes, which becomes
possible only in the right angle limit because of the $s$-rule.
The appearance of such
extra rational curves means a generation of
extra flat directions of the
$\mu=\infty$ theory. This will be important in the next section.
We consider only the case $N_f\geq N_c$.

We recall that for finite angle rotation and $\mu<\infty$
the projection of the fivebrane in the
Taub-NUT space (i.e. forgetting the $x^{0,1,2,3,7,8,9}$ direction)
remains the same as the starting point $\mu=0$,
provided we fix the $N=2$ scale
$\Lambda_{N=2}$. In terms of the rescaled variable
it is given by
\beq
x+\mu^{-N_c}\tily=v^{N_c}+\cdots\,.
\eeq
where $+\cdots$ are certain lower order terms in $v$.
Here we stress that this equation describes the whole curve
(projected onto the Taub-NUT space) including the rational curve
components as well as the infinite component(s).
The coefficients of the lower order terms
$+\cdots$ in $v$
are given by positive powers of $\Lambda_{N=2}$, and hence are
of negative powers in $\mu$ if $\Lambda_{N=1}$ is fixed.
Thus, the $\mu\to\infty$ limit of the projection of the curve
onto the Taub-NUT space
is given by
\beq
x=v^{N_c}.
\label{proj}
\eeq
In principle, this could be different from the projection
of the $\mu\to \infty$ curve because some components
can go away to infinity in the $w$ direction.
Later, we show that this does not happen in the case $N_f\geq N_c$
we are considering
and indeed the $w$ values of any rational curve components are
at finite values of $w$. As we will see, the $w$ values
are actually zero except for
the $N_f=N_c$ curves describing the non-baryonic branch root.

\begin{figure}[htb]
\begin{center}
\includegraphics[width=5.0in]{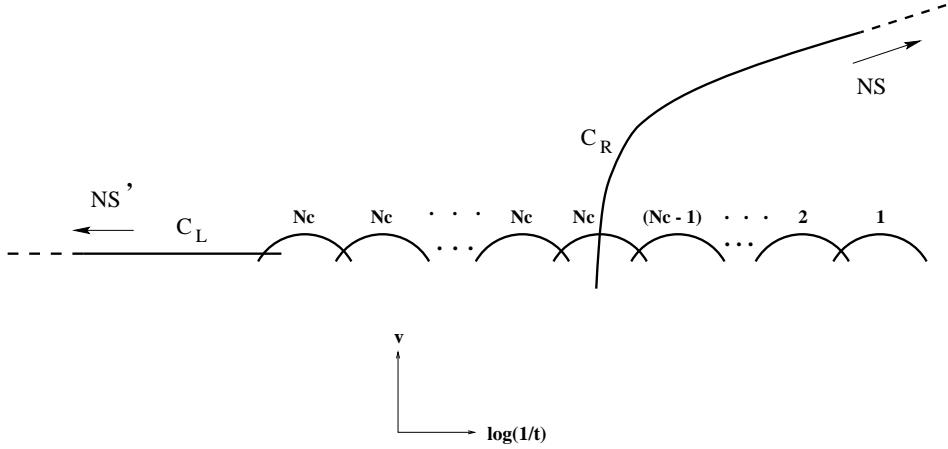}
\end{center}
\caption{The projection (\protect\ref{proj}) in the Taub-NUT space.}
\label{fig11}
\end{figure}

We first show how the projected curve (\ref{proj}) look like
in the Taub-NUT space. Recall that the Taub-NUT space
we are considering is the resolved
$A_{N_f-1}$ surface. It is covered by $N_f$ patches
with coordinates $(y_1,x_1),\ldots \ldots, \\ (y_{N_f},x_{N_f})$
which are projected to the $\tily$-$x$-$v$ space 
by\footnote{
We can obtain this from (\ref{projA}) in which we have put
$\Lambda_{N=2}=1$ for simplicity: We first recover $\Lambda_{N=2}$
in (\ref{projA}) just by replacing the expression for $y$ by
$y=\Lambda_{N=2}^{2N_c-N_f}y_i^ix_i^{i-1}$. Then, by definition
$\tily=\mu^{N_c}y=\Lambda_{N=1}^{3N_c-N_f}y_i^ix_i^{i-1}$, obtaining
(\ref{projA1}).
}
\beq
\begin{array}{l}
\tily=y_i^ix_i^{i-1}~,\\[0.2cm]
x=y_i^{N_f-i}x_i^{N_f+1-i}~,\\[0.2cm]
v=y_ix_i~,
\end{array}
\label{projA1}
\eeq
in which we have put $\Lambda_{N=1}=1$ for simplicity.
The equation (\ref{proj}) looks in the $i$-th patch as
\beq
y_i^{N_f-i}x_i^{N_f+1-i}
=y_i^{N_c}x_i^{N_c}\,.
\eeq
In the first $N_f-N_c$ patches ($i=1,\ldots,N_f-N_c$)
the RHS is of lower order both in
$y_i$ and $x_i$ than the LHS, while the LHS is of
lower order than RHS in the remaining patches
($i=N_f-N_c+1,\ldots,N_f$).
Recall that the equations $y_i=x_{i+1}=0$
define a rational curve $C_i$.
Thus, the curve includes as its components
the rational curve $C_i$ with multiplicity
$N_c$ for $1\leq i\leq
N_f-N_c$ and $N_f-i$ for $N_f-N_c+1\leq i\leq N_f-1$.
Namely, the multiplicities of
$C_1,\ldots,C_{N_f-1}$ are
\beq
\underbrace{N_c,N_c,\ldots, N_c}_{N_f-N_c},
N_c-1,\ldots,2,1\,.
\eeq
In addition to these rational curve components, there are
two components of infinite volume ---
the curve
described by $x_1^{N_c}=0$ (i.e. $y_1$-axis with multiplicity
$N_c$) in the first patch
and the curve described by
\hbox{$\scriptstyle y_i^{N_f-N_c-i}x_i^{N_f-N_c-i+1}=1$}.
As we will see, the actual curve projected to $x_1^{N_c}=0$ is
an $N_c$-fold cover $C_L$
extending in the $w$ direction,
while the curve $C_R$ projected to the latter is at $w=0$.
If $N_f>N_c$,
the projection of $C_R$
intersects with the rational curve $C_{N_f-N_c}$.
In summary, the projected curve (\ref{proj})
is depicted in figure \ref{fig11}.

Note that the numbers of rational or infinite curve components
have increased from any of the $N=2$ Higgs branch roots. 
This means that the infinite component of the finite $\mu$
curve degenerates in the $\mu\to\infty$ limit and
has turned into a union of rational curve components
and infinite component(s).
We now exhibit this explicitly, proving that (\ref{proj})
is indeed the projection of the $\mu\to\infty$ curve.

\bigskip
{\it Baryonic Branch Root}

\medskip
We recall that the curve of the baryonic branch root
consists of $N_fN_c-N_c^2$ rational curve components and two
infinite components $\tilC_L$ and $C_R$. The components other than
$\tilC_L$ will stay the same after rotation as well as in the
$\mu\to\infty$ limit.
So we only have to consider the $\mu\to\infty$ limit of $\tilC_L$.
The component $\tilC_L$ at finite $\mu$ given by
(\ref{bbmu}) is described in the $i$-th patch by
\beqa
y_i&=&\mu^{i-1}w^{N_c-i+1}\,,\\
x_i&=&\mu^{-i}w^{i-N_c}\,.
\eeqa
Let us see this in the limit $\mu\to\infty$.
If $i>N_c$, the first equation $y_iw^{i-N_c-1}=\mu^{i-1}$
can never be satisfied for finite $y_i$ or $w$, and thus the curve
disappears from this patch.
So let us consider the case $i\leq N_c$.
For $x_i$ to be finite, $x_i\sim 1$, $w$ must scale as
$\mu^iw^{N_c-i}\sim 1$. Then, $y_i=\mu^{-1}w/x_i\sim \mu^{-1}w$
behaves as $y_i^{N_c-i}\sim \mu^{-(N_c-i)}w^{N_c-i}\sim \mu^{-N_c}$.
This means that for every finite $x_i$, there are $N_c-i$ values
of $y_i$ approaching to zero. Thus,
we see that the limit includes the component $C_i$ with multiplicity
$N_c-i$. If we repeat the same thing by interchanging $x_i$ and $y_i$,
we see that the limit includes the component $C_{i-1}$ with
multiplicity $N_c-i+1$. The location of these components are
at $w=0$ because of the scaling relations ---
$\mu^iw^{N_c-i}\sim 1$ for $C_i$. In the first patch, there
is another component $C_L$ in which $w$ is arbitrary:
\beq
C_L\,\,\left\{
\begin{array}{l}
y_1=w^{N_c}\,,\\[0.1cm]
x_1=0\,.
\end{array}
\right.
\eeq

\begin{figure}[htb]
\begin{center}
\includegraphics[width=5in]{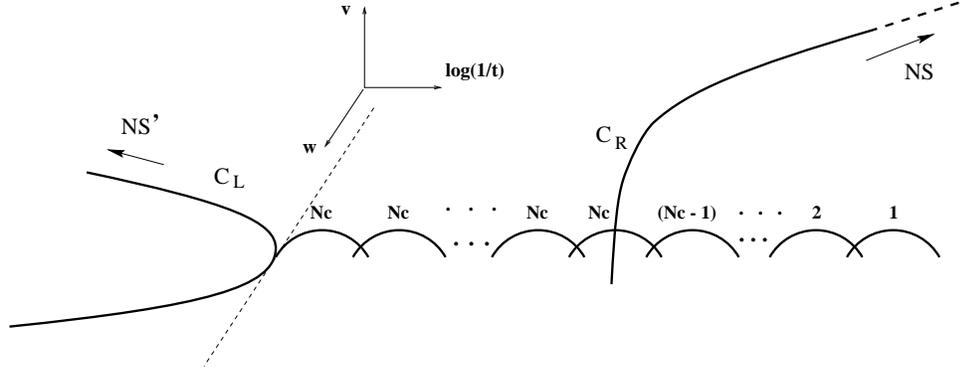}
\end{center}
\caption{The $\mu\to\infty$ limit of the baryonic branch root ($N_f>N_c$).}
\label{fig6}
\end{figure}

\begin{figure}[htb]
\begin{center}
\includegraphics[width=4.0in]{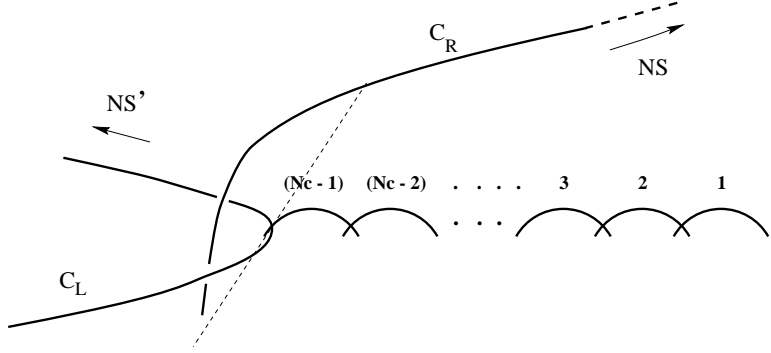}
\end{center}
\caption{The $\mu\to\infty$ limit of the baryonic branch root ($N_f=N_c$).}
\label{fig6b}
\end{figure}

In summary, the curve $\tilC_L$ degenerates in the $\mu\to\infty$
limit and consists of an infinite component $C_L$
and rational curves $C_1,C_2,\ldots,C_{N_c-1}$
of multiplicity
$N_c-1,N_c-2,\ldots,1$.
Together with the components
$C_1,C_2,\ldots,C_{N_f-1}$ and $C_R$
with multiplicity $1,2,\ldots,r_*-1,r_*,\ldots, r_*,r_*-1,\ldots,2,1$
and $1$ respectively (recall that $r_*=N_f-N_c$)
which we have been omitting for a while,
the total $\mu\to\infty$ curve consists of the components
$C_L,C_1,C_2,\ldots,C_{r_*},C_{r_*+1},\\ \ldots,C_{N_f-2},C_{N_f-1}$
and $C_R$ 
with the
multiplicities $1,N_c,N_c,\ldots,N_c,N_c-1,\break \ldots, 2, 1$ and $1$
respectively.
This is depicted in figure \ref{fig6} and figure \ref{fig6b}.


\bigskip
{\it Non-Baryonic Branch Roots}

\medskip
\noindent
\underline{$N_f=N_c$}

\begin{figure}[htb]
\begin{center}
\includegraphics[width=5.0in]{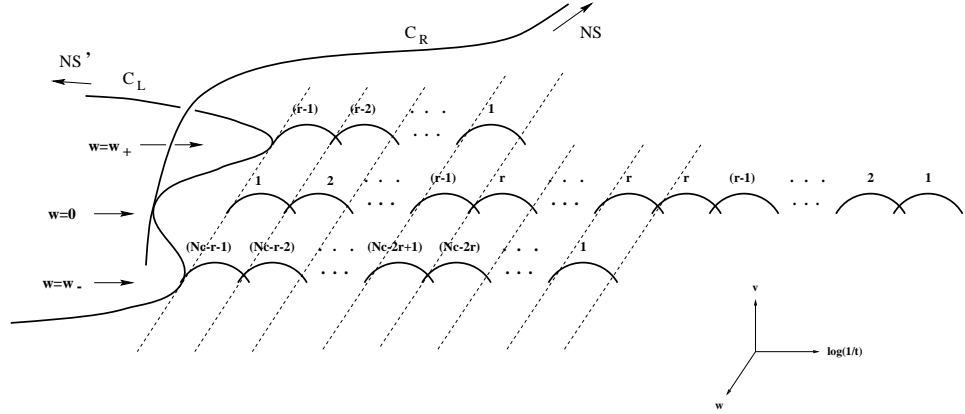}
\end{center}
\caption{The $\mu\to\infty$ limit of the non-Baryonic branch root
in the $N_f=N_c$ case.}
\label{fig12}
\end{figure}

Lets us look at the $\mu$ finite rotated curve of the
$r$-th non-baryonic branch root of the $N_f=N_c$ case
where $r=1,\ldots,[N_c/2]$.
The infinite component $C$ is described in the $i$-th patch by
\beqa
y_i&=&\mu^{i-1}w^{i-1}(w-w_+)^{r-i+1}(w-w_-)^{N_f-r-i+1}\\
x_i&=&\mu^{-i}w^{-i}(w-w_+)^{i-r}(w-w_-)^{i-(N_f-r)}\,.
\eeqa
Recall that for $N_f = N_c$ case, $w_+$ and $w_-$ have non-zero
distinct limits as $\mu \rightarrow \infty$. There are three ways
to scale $w$ in order to make $y_i$ and $x_i$ finite in the limit
$\mu \rightarrow \infty$:\\
\indent
(i) $w \rightarrow 0$ keeping $\mu w$ finite.\\
\indent
(ii) $w \rightarrow w_+$ keeping $\mu^{i-\epsilon}(w-w_+)
^{r-i+\epsilon}$ finite ($\epsilon =0,1$).\\
\indent
(iii) $w \rightarrow w_-$ keeping $\mu^{i-\epsilon}(w-w_-)
^{N_f-r-i+\epsilon}$ finite ($\epsilon =0,1$).\\
The scaling of type (i) is possible for every $i$, and the corresponding
limit is an infinite curve $C_R$ given by $y_i^i x_i^{i-1} = 1,
w=0$. The scaling of type (ii) is possible only for $i \leq r$.
For $\epsilon =0$, $x_i$ is finite and $y_i$ goes to zero as
$y_i^{r-i} \sim \mu^{-r}$, while for $\epsilon =1$, $y_i$ is finite
and $x_i$ goes to zero as $x_i^{r-i+1} \sim \mu^{-r}$. Thus,
from the type (ii) limit, we obtain the components
$C_1, C_2, ..., C_{r-1}$
at $w=w_+$ with multiplicity $(r-1), (r-2), ..., 1$ respectively.
The scaling of type (iii) is possible only for $i \leq N_f -r$. By the
similar argument as in type (ii), one can see that we obtain the
components $C_1, C_2, ..., C_{N_f-r-1}$ at $w=w_-$ with multiplicity
$(N_f-r-1), (N_f-r-2), ..., 1$ respectively. In the first patch
$i=1$, there is another component $C_L$ given by $y_1 =
(w-w_+)^r (w-w_-)^{N_f-r}$, $x_1=0$ which is infinite and parametrized
by $w$. Together with the components which already exist before taking
the limit, the $\mu \rightarrow \infty$ curve consists of the components
depicted in figure \ref{fig12}.

\noindent
\underline{$N_f > N_c$}

Finally, we consider the $\mu \rightarrow \infty$ limit of the
$r$-th non-baryonic branch root, where $r=0, 1, ..., [N_f/2]$,
$r \neq r_* = N_f - N_c$.

In this case, both $w_+$ and $w_-$ vanish in the limit $\mu
\rightarrow \infty$. More precisely, they scale as $w_\pm
\sim \mu^{-\beta}$ where $\beta = (N_f-N_c)/(2N_c-N_f)$.
There are three ways to scale $w$, keeping $y_i$ and
$x_i$ to be finite in the limit:\\
\indent
(i) $w \sim \mu^{-\alpha}$ where $\alpha < \beta$.\\
\indent
(ii) $w \sim \mu^{-\alpha}$ where $\alpha > \beta$.\\
\indent
(iii) $w \sim w_+ + \mu^{-\delta}$ where $\delta > \beta$.\\
\indent
(iv) $w \sim w_- + \mu^{-\delta}$ where $\delta > \beta$.\\
The type (ii) limit yields an infinite curve $C_R$ given by
$\scriptstyle y_i^{N_f-N_c-i} x_i^{N_f-N_c-i+1} = 1$. The type (i) limit
yields components $C_{i=1,...,N_c-1}$ with the multiplicity
of $C_i$ being $(N_c - i)$. The type (iii) limit yields
components $C_{i=r+1,...,r_*-1}$ with multiplicity $(i-r)$
if $r < r_*$, while it yields
components $C_{i=r_*+1,...,r-1}$ with multiplicity $(r-i)$
if $r>r_*$.
The type (iv) limit yields components
$C_{i=r_*+1,...,N_f-r-1}$
with multiplicity $(N_f-r-i)$. These components are all at
$w=0$. In the first patch, there is another component
$C_L$ given by $y_1=w^{N_c}, x_1=0$.

Together with the components which already exist before 
the rotation, the $\mu \rightarrow \infty$ limit of the curve
for the $r$-th non-baryonic branch root is the same as
the limit of the curve for the baryonic branch root which
is depicted in figure \ref{fig6}.

\section{More on the $N=1$ Moduli Space of Vacua}

In this section we will discuss the moduli
space of vacua of the $N=1$ theories
for finite as well as infinite adjoint mass $\mu$
including the baryonic degrees of freedom.

\subsection{M-theory Proof of the $s$-Rule}

We have seen in the previous sections that for finite adjoint mass 
the moduli space of vacua consists of  non-baryonic
branches parametrized by an integer $r$ and having
complex dimension $2r(N_f-r)$ and a baryonic branch of complex dimension
$2N_cN_f - 2(N_c^2-1)$ which exists when $N_f \geq N_c$.
In the limit $\mu \rightarrow \infty$ these different branches become
submanifolds
of the Higgs branch of $N=1$ SQCD.
The complex dimension of the Higgs branch of $N=1$ SQCD is 
$2N_cN_f - (N_c^2-1)$ which means that in the limit 
$\mu \rightarrow \infty$ an extra $N_c^2 -1$ degrees of freedom become 
massless. We will start by showing how this can be seen in the M-theory
fivebrane picture.

\begin{figure}[htb]
\begin{center}
\includegraphics[width=5.0in]{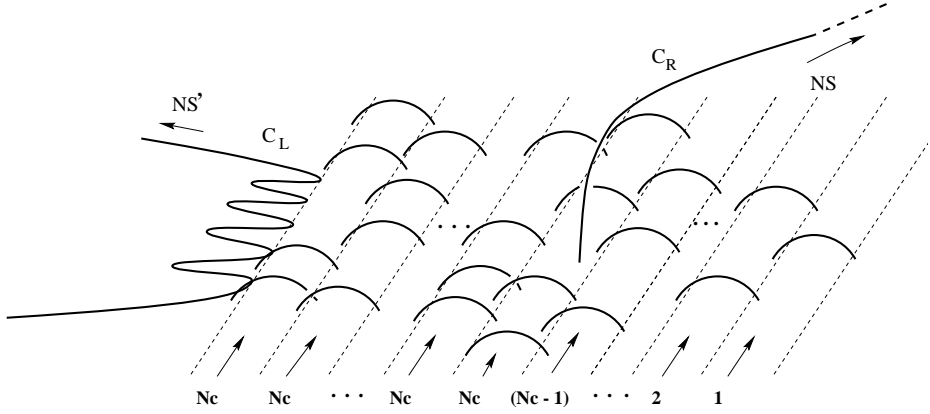}
\end{center}
\caption{M-Theory description of the Higgs branch.}
\label{fig7}
\end{figure}

As in previous cases we have to describe the curve
in the resolved $A_{N_f-1}$ surface.
This is depicted in figure \ref{fig7}.
The left component $C_L$
of the curve describing the NS' 5-brane is given by
\beq
\tilt=w^{N_c},~~~~~
v=0
\label{l}
\comma
\eeq
while the right component $C_R$
of the curve describing the NS 5-brane is given by
\beq
\tilt=\Lambda_{N=1}^{3N_c-N_f}v^{N_f-N_c},~~~~~
w=0
\label{right}
\stop
\eeq
Possible deformations of the left component (\ref{l}) are
\beq
\tilt=w^{N_c}+s_1w^{N_c-1} +... + s_{N_c},~~~~~
v=0
\label{ld}
\stop
\eeq
It is not possible to deform the right component (\ref{right}).
This follows from the invariance under the action of
$U(1)_{4,5}$ and $U(1)_{8,9}$,
where the charges are given in (\ref{list1}).
Since there is no order parameter carrying only $U(1)_{4,5}$
charges this symmetry cannot be broken and the equation
(\ref{right}) is fixed, while 
the vev of the meson field can break the $U(1)_{8,9}$
symmetry and corresponds to the deformation
(\ref{ld}).
The fact that there are no allowed deformations 
of (\ref{right})
can also be deduced in the case $N_f\geq 1$ by the
following argument.
Suppose that we try to deform it by
adding other monomials in $v$. This implies that there will 
be non-zero values of $v$
for which $\tilt=0$.
Recall, however, that the $(\tilt,v)$ coordinates are of
the Taub-NUT space 
\beq
\tilt x = \Lambda_{N=1}^{3N_c-N_f} v^{N_f}
\label{ale}
\comma
\eeq
which excludes $\tilt=0$ for non-zero values of $v$.
Therefore deformations of (\ref{right}) are not allowed.
This in fact provides a {\it proof } of the $s$-rule.

\begin{figure}[htb]
\begin{center}
\includegraphics[width=3.8in]{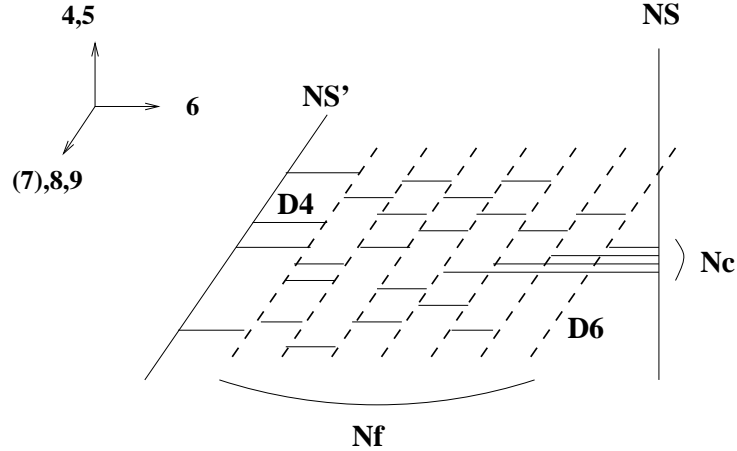}
\end{center}
\caption{Type IIA description of the Higgs branch.}
\label{fig4}
\end{figure}

\begin{figure}[htb]
\begin{center}
\includegraphics[width=3.3in]{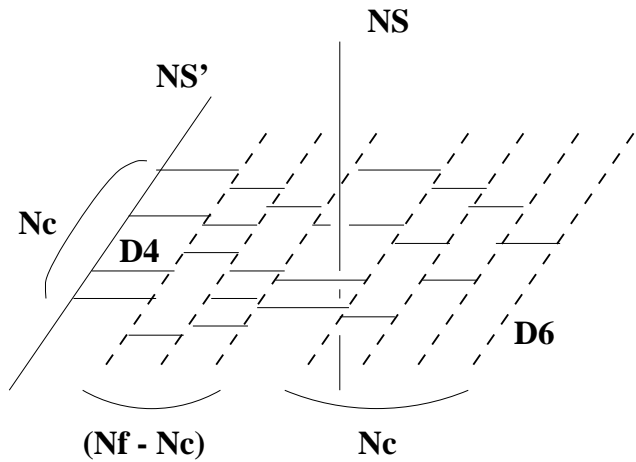}
\end{center}
\caption{Alternative type IIA description of the Higgs branch.}
\label{fig5}
\end{figure}

To see that this indeed corresponds to the $s$-rule, let us
compare IIA and M-theory brane configurations. 
The Higgs branch of $N=1$ SQCD is depicted in figure \ref{fig4}.
For comparison with the M-theory description of the 
Higgs branch it is useful to
move the right NS 5-brane in figure \ref{fig4} 
and pass $N_c$ D6-branes. The configuration that
we end with is
plotted in figure \ref{fig5}. 

The $\CP^1$ components in figure \ref{fig7} correspond to the D4-branes
suspended between the D6-branes in the IIA picture of figure \ref{fig5}.
The $N_c$ complex
moduli associated with the $N_c$ D4-branes suspended between 
the NS' 5-brane and the D6-brane
to its left are seen  in the M-theory fivebrane
framework as deformations of the left part of the curve in
figure \ref{fig7}  which is the part of the curve  describing the NS' 5-brane.
The existence of allowed deformations of the  
 left part of the curve in
figure  \ref{fig7} that describes the NS' 5-brane and the
non-existence
of allowed 
deformations of the  
 right part of the curve in
figure \ref{fig7}  that describes  the NS 5-brane is
from the IIA viewpoint the $s$-rule which has been 
suggested empirically in \cite{hw}.

Here we have shown that (\ref{ld}) are the only permissible
deformations of the left and the right curves, consistent
with the asymptotic conditions and the $U(1)$ R-symmetry.
To show that
all these deformations actually
correspond to vevs of fields in the $N=1$ theory, we need to compute
the K\"ahler metric for these deformations and show that
it is regular and non-degenerate. This issue is
currently under investigation \cite{new}.

A simple counting of the number of D4-branes suspended between 
the D6-branes and between the
D6-branes and the NS' 5-brane gives  
$(2N_cN_f - N_c^2)$ as the complex dimension of the $N=1$ theory.
As in the $N=2$ case the type IIA brane counting results in the dimension
of a $U(N_c)$ gauge group instead of an $SU(N_c)$ gauge group, missing
one complex dimensions in the moduli space. It is possible that
this missing dimension corresponds to the relative locations of the
NS and NS' 5-branes in the $x^7$-direction and its superpartner.
In order to fully establish this, we need to compute the K\"ahler metric
for this direction.

\subsection{The Baryonic Degrees of Freedom}

In section 4 we showed how the meson matrix is realized
in the M-theory fivebrane.
For $N_f \geq N_c$ there are also baryonic degrees of freedom.
In the following we will show how
the latter are realized in the M-theory fivebrane.
We will derive the equations (\ref{mbb}), (\ref{ncf}) 
 and (\ref{final}) of section 2 from the fivebrane.

\begin{figure}[htb]
\begin{center}
\includegraphics[width=2.0in]{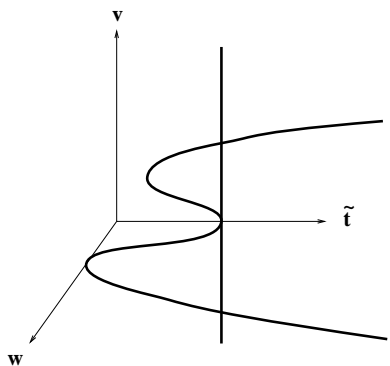}
\end{center}
\caption{The two components of (\protect\ref{nbbnbb}) for $N_c=N_f=2$.}
\label{fig8}
\end{figure}

We start with the case $N_f=N_c$. The vev for the baryon
operators is zero on the 
non-baryonic branches and is 
non zero  on the baryonic branch. 
In order to illustrate the 
  difference between the non-baryonic and baryonic
branches let us consider the 
 $r$-th  non-baryonic branch with $r=\frac{N_f}{2}$.
In the limit $\mu \rightarrow \infty$
 it is described by (\ref{nf=nccurve}) 
\beq
\left\{
\begin{array}{l}
\tilt = (w^2+\Lambda_{N=1}^4)^{\frac{N_c}{2}}\,,\\
v=0\,,
\end{array}
\right.
\qquad
\left\{
\begin{array}{l}
\tilt = \Lambda_{N=1}^{2N_c}\,,\\
w=0\,.
\end{array}
\right.
\label{nbbnbb}
\eeq
In figure \ref{fig8} we plot for simplicity the two components
of the curve (\ref{nbbnbb}) 
in the case $N_c=N_f=2$.
The above non-baryonic branch is part of the $N=1$ SQCD Higgs
branch where $\tilde{B}B=0$.

\begin{figure}[htb]
\begin{center}
\includegraphics[width=2.5in]{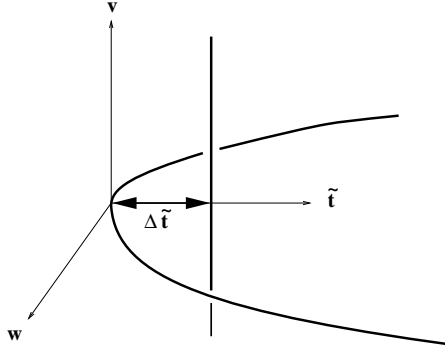}
\end{center}
\caption{The two components of (\protect\ref{bbbb}).}
\label{fig9}
\end{figure}

For comparison consider now the baryonic branch in the limit   
$\mu \rightarrow \infty$.
It is given by (\ref{bb})
\beq
\left\{
\begin{array}{l}
\tilt = w^{N_c}\,,\\
v=0\,,
\end{array}
\right.
\qquad
\left\{
\begin{array}{l}
\tilt = \Lambda_{N=1}^{2N_c}\,,\\
w=0\,.\\
\end{array}
\right.
\label{bbbb}
\eeq
In figure \ref{fig9} we plot the two components of (\ref{bbbb}).
The baryonic branch is part of the $N=1$ SQCD Higgs
branch where $\tilde{B}B \neq 0$.

Comparing figures \ref{fig8} and \ref{fig9} we see
that while in figure \ref{fig8} 
the two branches intersect,
in figure \ref{fig9} they do not.
The distance between the two branches at $w=0$ is a candidate
for $\tilde{B}B$, when the vev of the meson matrix vanishes. 
As a first check we have to see that 
it carries the  $U(1)_R$ and $U(1)_A$ charges of $\tilde{B}B$.
Using (\ref{list1}) and the relation
between the $U(1)_R$ and $U(1)_A$ charges and 
the $U(1)_{45}$ and $U(1)_{89}$ charges:
\beqar
U(1)_R &=& \frac{N_c}{N_f} U(1)_{45} + \frac{N_f-N_c}{N_f} U(1)_{89}
\nonumber\\
U(1)_A &=& -U(1)_{45} + U(1)_{89} 
\comma
\eeqar
we see that $\tilde{t}$ carries the charges $(U(1)_R,U(1)_A)
= (2N_c\frac{N_f-N_c}{N_f},
2N_c)$ which are the correct charges that  
$\tilde{B}B$ carries.

\begin{figure}[htb]
\begin{center}
\includegraphics[width=2.5in]{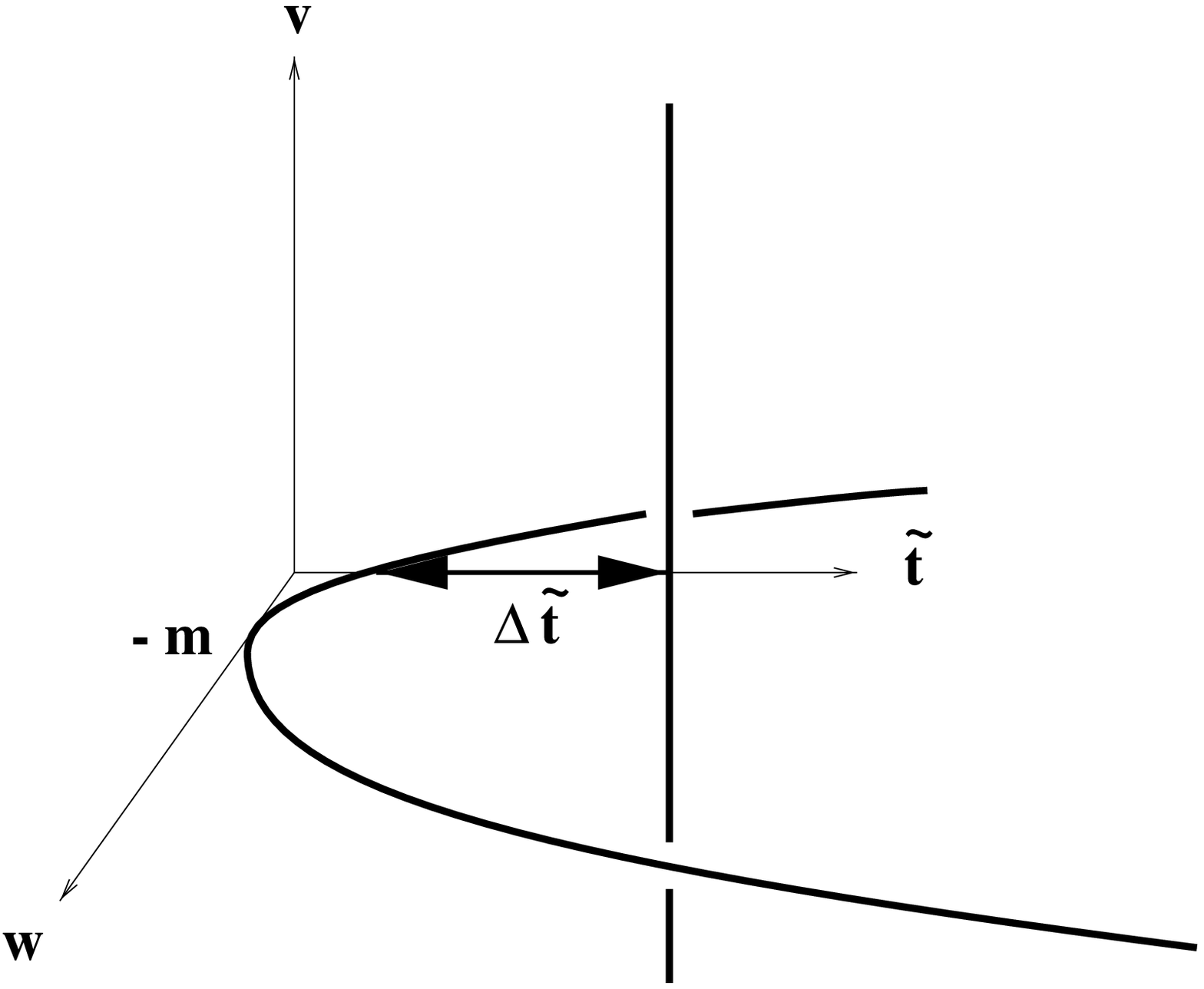}
\end{center}
\caption{Shifting $w$ in  (\protect\ref{bbbb}).}
\label{fig10}
\end{figure}

A non-trivial check is to verify that 
when we shift the second branch of (\ref{bbbb}) by shifting $w$
as in figure \ref{fig10}  which means  giving a 
vev for the meson, the values of the distance and the shift have
to satisfy the relation
(\ref{mbb}).

Indeed  we see that the distance $\Delta \tilt$
between
the two branches at $w=0$ is
\beq
 m^{N_c}-\Delta \tilt = \Lambda_{N=1}^{2N_c}
 \comma
 \label{tt}
 \eeq
 which is  identical to (\ref{mbb}) if we identify the distance 
$\Delta \tilt$ as $\tilde{B}B$.
 Note also, that the brane equation (\ref{tt}) captures the fact
  that in $N=1$ SQCD with $N_f=N_c$
the classical moduli space of vacua is modified quantum mechanically.

Consider next the case $N_f=N_c+1$, and let $\mu$ be finite.
The two components of the curve
corresponding to the baryonic branch are given by (\ref{bbmu}).
As in the case $N_f=N_c$ we shift $w \rightarrow w+m$ and get 
\beq
\left\{
\begin{array}{l}
\tilt=(w+m)^{N_c}\,,\\
v=\mu^{-1}(w+m)\,,
\end{array}
\right.
\qquad
\left\{
\begin{array}{l}
\tilt = \Lambda_{N=1}^{3N_c-N_f}v\,,\\
w=0\,.
\end{array}
\right.
\label{bb1}
\eeq
It is easy to compute the distance $\Delta \tilt$
between the two components of the curve (\ref{bb1}) at
$w=0$ and we get
\beq
 m^{N_c}-\Delta \tilt = \Lambda_{N=1}^{3N_c-N_f}\frac{m}{\mu}
 \comma
 \label{ttt}
 \eeq
which is  identical to (\ref{ncf}) if we identify the distance 
$\Delta \tilt$ as $\tilde{B}B$.
Note however that there is a difference between
the $N_f=N_c+1$ and $N_f=N_c$ cases.
In the former  $\tilde{B}B$ is a matrix
while in the latter it has a single component.
Clearly the distance between the two parts
of the baryonic branch curve cannot provide
us with the information on the full  $\tilde{B}B$  matrix. Equations
(\ref{ttt}) and (\ref{ncf}) describe
only part of the possible vev's for the baryons. 
In the limit $\mu \rightarrow \infty$ the RHS of
(\ref{ttt}) vanishes and we get the
description of the moduli space of vacua of $N=1$ SQCD with $N_f=N_c+1$.
In this case,
the classical moduli space of vacua is not modified quantum mechanically.

Finally, let  $N_f > N_c+1$, and keep $\mu$ finite.
In this case there was an ambiguity in the field theory
analysis in finding the extrema of
the superpotential in section 2.
The two components of the curve
corresponding to the baryonic branch read now
\beq
\left\{
\begin{array}{l}
\tilt = (w+m)^{N_c}\,,\\
v=\mu^{-1}(w+m)\,,
\end{array}
\right.
\qquad
\left\{
\begin{array}{l}
\tilt = \Lambda_{N=1}^{3N_c-N_f}v^{N_f-N_c}\,,\\
w=0\, ,
\end{array}
\right.
\label{bb2}
\eeq
where we shifted $w \rightarrow w+m$.
The distance $\Delta \tilt$
between the two components of the curve (\ref{bb2}) at
$w=0$  satisfies
\beq
 m^{N_c}-\Delta \tilt = \mu^{N_c-N_f}m^{N_f-N_c}\Lambda_{N=1}^{3N_c-N_f}
 \stop
 \label{tttt}
 \eeq
 This relation is identical to (\ref{final}) which we derived
 by field theory means with one choice of approaching the region
in moduli space where the
 determinant of the meson matrix vanishes. 
 Unlike the field theory analysis, there is no ambiguity
in the brane framework in deriving
 (\ref{bb2}). This is not surprising since we expect the brane
 to provide us with the good coordinates on the moduli space of vacua. The 
 ambiguity that we encountered in section 2  can be traced
 to the fact that ${\det M}=0$ defines a moduli space and
we need information about the 
 good coordinates near that region in order to approach it.
 As in the previous case,
 in the limit $\mu \rightarrow \infty$ the RHS of (\ref{tttt})
vanishes and we get
 the correct description of the moduli space of
vacua of $N=1$ SQCD with $N_f > N_c+1$, 
 where
the classical moduli space  is not modified quantum mechanically.

 To summarize: We showed in this section how 
the baryonic degrees of freedom (up to the chiral rotation)
are realized in M-theory fivebrane. 
In particular we found a complete agreement
 for both finite and
  infinite values of the adjoint mass between
the field theory results of section 2 and the
  fivebrane description of the moduli space of vacua.

\section{Beyond Holomorphy --- K\"ahler Potential}

We have shown that the configuration of the fivebrane
encodes strong coupling physics of the $N=1$ theory, 
in particular the Affleck-Dine-Seiberg superpotential 
for $N_c > N_f$ and the holomorphic structure of the 
moduli space for $N_c \leq N_f$. In the field theory
approach, these non-perturbative results were
obtained using the holomorphy argument. In order to fully 
understand the low energy dynamics, however, we also need 
to determine the K\"ahler potential. In the $N=1$ theory, 
the K\"ahler potential is independent of the superpotential 
and the holomorphy argument is not sufficient to specify it.  

We expect, in this regard, the fivebrane approach to be
more powerful than the standard field theory method. 
The eleven-dimensional Planck scale $l_{11}$, 
the radius $R$ of the eleventh dimensional circle, the type 
IIA string coupling constant $g_s$ and the string scale $l_s$ 
are related as
\beq
     l_{11} \sim g_s^{1/3} l_s,~~~ R \sim g_s l_s.
\label{last1}
\eeq
On the other hand, the gauge coupling constant $g_{gauge}$ 
of $N=1$ theory that arises from the web of branes scales 
\cite{hw} as 
\beq
    g_{gauge}^2 \sim \frac{g_s l_s}{L_{brane}},
\label{last3}
\eeq
where $L_{brane}$ is the distance between the NS and NS'
5-branes.  
Therefore
\beq
   g_{gauge} \sim \left( \frac{R}{L_{brane}} \right)^{1/2}.
\eeq
This means that we can take the limit $R/l_{11}, L_{brane}/l_{11}
\rightarrow \infty$ 
while keeping $g_{gauge}$ finite, and  the low energy 
effective theory of M-theory, namely the eleven-dimensional 
supergravity, should give an exact description of 
$N=1$ theory. Low energy degrees of freedom of the fivebrane are
its deformation in the spacetime and the chiral 
anti-symmetric tensor field on the brane. Thus, we 
can directly read off their kinetic 
terms from the fivebrane action \cite{sch}
and find the K\"ahler metric.

One of the interesting questions concerning the K\"ahler potential
is its behavior at the origin of the moduli space for 
$N_c < N_f$. If one can show, for example 
that the K\"ahler metric is regular at the origin 
for $N_f = N_c+1$, it gives a direct confirmation of the claim 
(based on the 't Hooft anomaly matching condition) that both 
the mesons and the baryons are massless degrees of freedom
there. Since the superpotential for $N_f = N_c + 1$ is given by
\beq
    W = \Lambda^{1-2N_c} \left(  \tilde{B}M B
     - {\det M} \right),
\eeq
if the metric $g_{i\bar{j}}$ is regular, 
the potential for the scalar fields 
\beq
     V = g^{i\bar{j}} \partial_i W \bar{\partial}_{\bar{j}}
        \bar{W}
\eeq
vanishes quartically and is flat at the origin. We hope
to discuss more on this subject in our future publication \cite{new}.  

\advance\baselineskip by 0pt plus 0.8pt
\advance\parskip by 0pt plus 2pt

\section*{Note added}
After completing this work we were informed of \cite{wittennew}
where related issues are studied.

\section*{Acknowledgements}

We would like to thank D. Kutasov, H. Murayama, R. Plesser,
C. Vafa and E. Witten 
for valuable 
discussions. K.H. would like to thank Institute for Advanced Study and
Rutgers Physics Department, H.O. would
like to thank Harvard and Rutgers Physics Departments
and Y.O. would
like to thank the Weizmann Institute Physics Department, for hospitality.

This research is supported in part by 
NSF grant PHY-95-14797 and DOE grant DE-AC03-76SF00098.

\end{document}